\documentclass{article}

\usepackage{arxiv}

\usepackage{amsmath}
\usepackage[utf8]{inputenc} 
\usepackage[T1]{fontenc}    
\usepackage{hyperref}       
\usepackage{url}            
\usepackage{booktabs}       
\usepackage{amsfonts}       
\usepackage{nicefrac}       
\usepackage{microtype}      
\usepackage{cleveref}       
\usepackage{lipsum}         
\usepackage{graphicx}
\usepackage{natbib}
\usepackage{doi}
\usepackage{multirow}
\usepackage{amssymb}
\usepackage{algorithm,algorithmic}

\title{Energy Consumption of GEO-to-ground Beaconless Link Acquisition Against Random Vibration with Coherent Detection}

\newif\ifuniqueAffiliation
\uniqueAffiliationtrue

\ifuniqueAffiliation 
\author{ \href{https://orcid.org/0000-0002-2471-1574}{\includegraphics[scale=0.06]{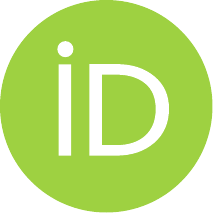}\hspace{1mm}Sen Yang} \\
	School of Astronautics and Aeronautics\\
	University of Electronic Science and Technology of China\\
	Chengdu, China 611731 \\
	\texttt{yang\_jansen@163.com} \\
	\And
	\href{https://orcid.org/0000-0000-0000-0000}{\includegraphics[scale=0.06]{orcid.pdf}\hspace{1mm}Xiaofeng Li} \\
	School of Astronautics and Aeronautics\\
	University of Electronic Science and Technology of China\\
	Chengdu, China 611731 \\
	\texttt{lxf3203433@uestc.edu.cn} \\
}
\else
\usepackage{authblk}

\newbox{\orcid}\sbox{\orcid}{\includegraphics[scale=0.06]{orcid.pdf}} 
\author[1]{%
	\href{https://orcid.org/0000-0002-2471-1574}{\usebox{\orcid}\hspace{1mm}Sen Yang\thanks{\texttt{yang\_jansen@163.com}}}%
}
\author[1]{%
	\href{https://orcid.org/0000-0000-0000-0000}{\usebox{\orcid}\hspace{1mm}Xiaofeng Li\thanks{\texttt{lxf3203433@uestc.edu.cn}}}%
}
\affil[1]{School of Astronautics and Aeronautics, University of Electronic Science and Technology of China, Chengdu, China 611731}
\fi

\renewcommand{\shorttitle}{\textit{arXiv} Template}

\begin{document}
\maketitle

\begin{abstract}
The GEO satellite maintains good synchronization with the ground, reducing the priority of acquisition time in the establishment of the optical link. Whereas energy is an important resource for the satellite to execute space missions, the consumption during the acquisition process rises to the primary optimization objective. However, no previous studies have addressed this issue. Motivated by this gap, this paper first model the relationship between the transmitted power and the received SNR in the coherent detection system, with the corresponding single-field acquisition probability, the acquisition time is then calculated, and the closed-form expression of the multi-field acquisition energy consumption is further derived in scan-stare mode. Then for dual-scan technique, through the induction of the probability density function of acquisition energy, it is transformed into the equivalent form of scan-stare, thereby acquiring acquisition energy. Subsequently, optimizations are performed on these two modes. The above theoretical derivations are verified through Monte Carlo simulations. Consequently, the acquisition energy of dual-scan is lower than that of scan-stare, with the acquisition time being about half of the latter, making it a more efficient technique. Notably, the optimum beam divergence angle is the minimum that the laser can modulate, and the beaconless acquisition energy is only 6\% of that with the beacon, indicating that the beaconless is a better strategy for optical link acquisition with the goal of energy consumption optimization.
\end{abstract}

\keywords{Optical Link Acquisition \and Coherent Detection \and Platform Vibration \and Multi-field Energy Consumption \and Optimizations}

The incorporation of big data in satellite communication has raised bandwidth demands of over 10 Gbps. Microwave communication systems are no longer sustainable in high-speed communication due to their low bandwidth magnitude and slow modulation rate\cite{toyoshima2007comparison}. In contrast, free-space optics communication (FSOC) can effectively be implemented as it provides several benefits, including a vast bandwidth, license-free spectrum, low power consumption, and minimal space requirements\cite{toyoshima2005trends}. As a result, numerous satellite-to-ground laser communication verification missions have gained significant success to date. From the LEO-to-ground laser communication experiment of STRV-2 module in 2000\cite{kim2001lessons}. Then the first high data rate (5.625 Gbps) bidirectional link using 1064nm coherent waveform between the NFIRE satellite and the optical ground station in 2010\cite{fields20115}. To current routine high-performance inter-satellite links between the Alphasat on GEO and the Sentinel-1A/1B on LEO\cite{benzi2016optical}. And the future EDRS provides global coverage for fast data transmission between network nodes and offers long-distance secure point-to-point communication\cite{hauschildt2019global}.

\begin{figure}[t!]
	\centering\includegraphics[width=1\columnwidth]{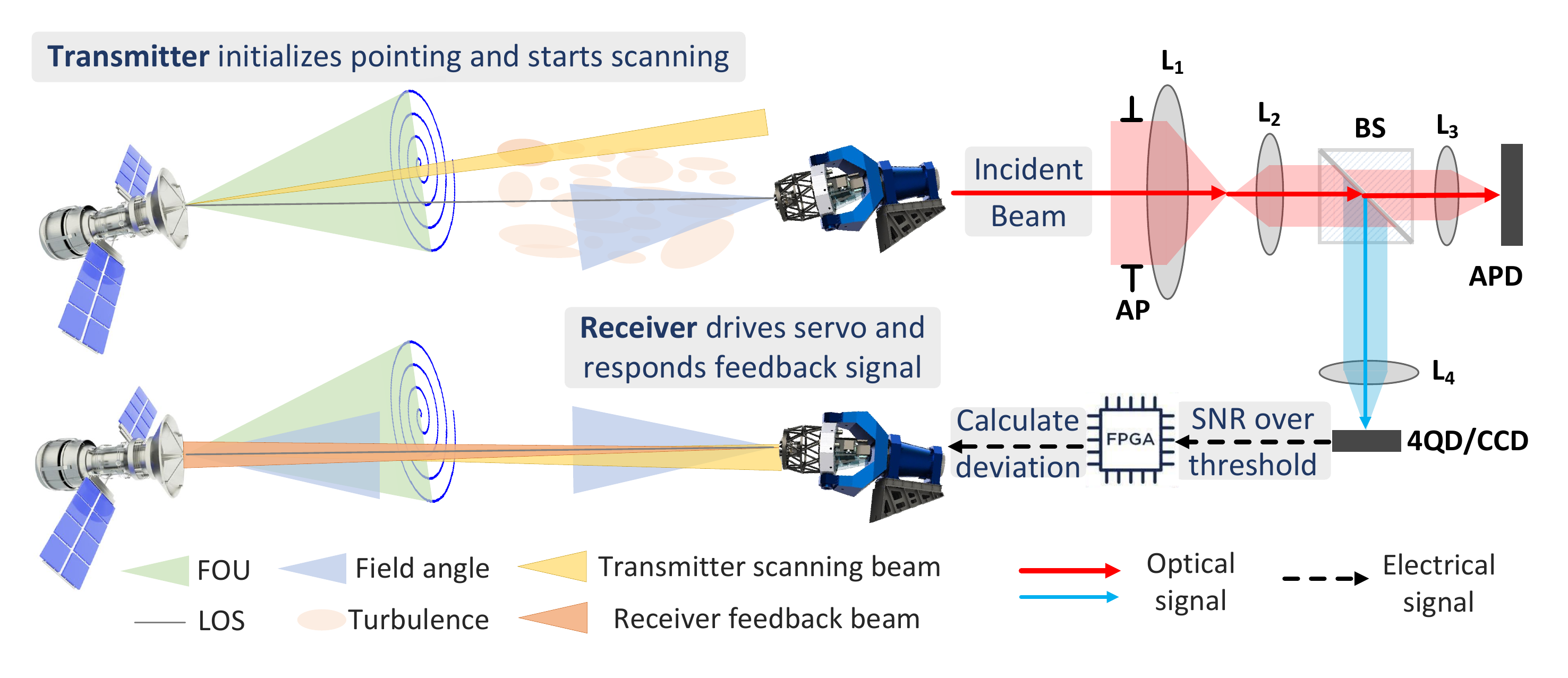}
	\caption{The diagram of beaconless acquisition. There is an error between the initial pointing and the LoS. The green conical covers the LoS so that the beam can propagate to the receiver antenna by scanning. Analogously, the field of view in blue conical covers LoS so that the received photons of the signal beam can fall on the photoelectric sensor. The incident beam passes through an AP and is then expanded by the telescope system $L_1-L_2$. It subsequently passes through a BS and most of it is focused on an APD through $L_3$ for signal processing. The remaining small portion of the beam is detected by a 4QD or CCD through $L_4$ for spot deviation measurement. $L_1,L_2,L_3,L_4$, lens; AP, aperture; BS, beam splitter; APD, avalanche photo diode; 4QD, four-quadrant detector; CCD, charge-coupled device photodetector.}
	\label{fig:1}
\end{figure}

The crucial step for establishing FSOC is optical link acquisition\cite{young1986pointing}, which has also been employed in precision measurement missions, such as GRACE FO\cite{wuchenich2014laser} for Earth gravitational field measurement, as well as LISA\cite{barausse2020prospects} and TAIJI program\cite{luo2021taiji} for space gravitational wave observation. Herein, we provide a typical optical link acquisition process to refine the background information\cite{guelman2004acquisition} . The transmitted and the received terminals on both sides of the optical link are defined as T1 and T2, respectively. Firstly, T1 and T2 utilize an ephemeris table for initial pointing. Due to differences in attitude and ephemeris, as well as thermal deformation and other factors, there is a certain angular deviation between the initial pointing and the line-of-sight (LoS), which is distributed randomly\cite{toyoshima2002optimum}, resulting in uncertainty for beam pointing. Then T1 performs a scanning with the Archimedes spiral\cite{steinhaus1999mathematical} to cover the field of uncertainty (FOU) of T2, while T2 maintains a staring posture. When T2 receives a laser signal with sufficient intensity, the distortion of spot on photodetector is corrected with adaptive optics\cite{yang2022derivation,yang2022iterative}, the deviation of the spot is calculated\cite{fu2021virtual,qiu2021active} and the pointing is adjusted slightly so that the spot moves to the center of the photodetector. Subsequently, T2 feeds back the optical signal. After the response of the photodetector of T1 is triggered, T1 stops scanning and the acquisition is completed The diagram is illustrated in Fig. \ref{fig:1}.

However, the advantage of good security brought by the smaller beamwidth has been proven to be a major drawback in the establishment of an optical link. Some studies adopted the beacon acquisition mode, where a wide beacon beam with sufficient peak power is used for scanning initially, followed by switching to a narrow beam for precise tracking when the two terminals are basically aligned. \cite{yu2017theoretical} studies the constraint boundary conditions for acquiring the beacon from the perspective of theory and experiment. A multi-parameter influenced pointing jitter error structure for the low orbit communication experimental satellite system has been established in\cite{hu2022multi}. However, in addition to the signal light laser, this system also requires a beacon laser device, which is not conducive to the miniaturization design of the system.

Ref.\cite{hindman2004beaconless} proposed a beaconless acquisition method that achieves scanning acquisition and LoS correction solely with the use of a narrow signal beam. Compared to the classical beacon, the beaconless method simplifies the terminal structure and reduces power requirements, while maintaining performance levels. However, the acquisition process poses significant challenges due to the narrow beam divergence angle and random vibration disturbance\cite{ho2007pointing}. To address these challenges, analytic expressions and optimizations for multi-scan average acquisition time were presented in\cite{li2011analytical}. In addition, an approximate mathematical model investigating the influence of Gaussian random vibration on the acquisition probability was proposed in\cite{friederichs2016vibration}. Ref.\cite{ma2021satellite} derived an approximate analytical expression for the scan loss probability with platform vibration, whose influence on acquisition time was analyzed under both single-field and multi-field scanning scenarios. Furthermore, Ref.\cite{hechenblaikner2021analysis} assessed the probability of failing to acquire a link due to beam jitter and derived a simple analytical model that allowed determining the maximum tolerable jitter for a given beam overlap and required probability of success. These models assumed that the scanning beam was detected with a sufficiently high signal-to-noise ratio (SNR). However, the actual transmitted power is limited and will also be affected by turbulence\cite{kaushal2016optical} and link losses, resulting in a significant attenuation of the power incident on the photodetector of the receiver. Moreover, there is noise output caused by dark current or other factors on the sensors\cite{li2015limited}, making the receiver SNR an important parameter that cannot be ignored.

The GEO satellite maintains good synchronization with the ground, and its satellite-to-ground communication window is wide. Hence, the establishment of the GEO-to-ground optical link does not impose stringent requirements on acquisition time. Rather, optimizing energy consumption, a crucial resource for satellites to execute space missions\cite{chin2018energy,cui2023energy}, should be the primary objective in the acquisition process. Nevertheless, to the best of our knowledge, there is no work has yet studied on it. Spurred by this gap, this paper provides detailed studies and discussions on the modeling and optimization of energy consumption for GEO-to-ground optical link acquisition. Considering the combined effects of platform vibration following the Rice distribution and atmospheric turbulence obeying the Gamma-Gamma distribution, we first model the transmitted power under a given average SNR at the receiver. Then, with the corresponding single-field acquisition probability, the probability density function (PDF) of multi-field acquisition time is derived. By integrating these components, the closed-form expression of the acquisition energy consumption is obtained in the scan-stare mode. Additionally, we conduct parameter optimizations for spiral pitch, beam divergence angle, coverage factor, and FOU. For the dual-scan mode, it is treated as two independent scan-stare processes. Subsequently, the PDF of multi-field acquisition energy consumption is derived and generalized to an equivalent scan-stare form, which enables to derive the expectation of multi-field acquisition energy consumption, and carry out optimizations. Finally, the Monte Carlo (MC) simulations also including the analysis of platform vibration and scanning speed are performed. 

The structure of the remaining sections of this article is as follows: Chapter 2 presents the model for transmitted power. Chapter 3 computes the energy consumption expectation for single-field acquisition. Building on this, Chapter 4 derives the energy consumption expectation for multi-field acquisition in both scan-stare and dual-scan modes, and performs optimization of acquisition parameters. Chapter 5 validates the theoretical derivations and optimization conclusions through MC simulations. Lastly, Chapter 6 offers a summary of the paper.

\section{Power Model}
The average SNR of FSOC system is\cite{andrews2005laser}:
\begin{equation}
	\bar{Q}={\left\langle i_{s}^{2} \right\rangle }/{\left\langle i_{n}^{2} \right\rangle }
	\label{eq:1}
\end{equation}
where $i_{s}$ is signal current. $i_{n}$ is noise current, which is additive Gaussian white noise with zero mean and $\sigma _{n}^{2}={{N}_{0}}$ variance. Due to the long transmission distance of GEO-to-ground laser communication, it is difficult to meet the demand of high communication rate by direct detection\cite{popoola2012spatial} under the condition of limited power. In contrast, the coherent detection offers advantages such as high receiver sensitivity, high communication rates, and strong resistance to background light interference. Consequently, in the coherent detection system, the photocurrent output by the balanced detector is\cite{peppas2015free}:
\begin{equation}
	{{i}_{s}}=2\sqrt{{{P}_{r}}{{P}_{L}}}{{R}_{r}}\cos (\Delta w\cdot t+\Delta \phi )
	\label{eq:2}
\end{equation}
where ${{P}_{r}}={{P}_{t}}{{h}_{t}}{{h}_{c}}$, $P_{t}$ and $P_{r}$ are the transmitter and receiver powers, respectively, $h_t$ represents the transmission gain with vibration, $h_c$ represents the turbulence attenuation, the two are independent\cite{jurado2012impact}. ${{R}_{r}}$ is the photoelectric response efficiency, ${{P}_{L}}$ is the local oscillator power, $\Delta w$ is the frequency difference between the signal and the local oscillator light, and $\Delta \phi $ is the phase difference between the signal and the local oscillator light. Therefore, the Eq. (\ref{eq:1}) is expressed as:
\begin{equation}
	\bar{Q}={2{{P}_{t}}{{P}_{L}}R_{r}^{2}E\left[ {{h}_{t}} \right]E\left[ {{h}_{c}} \right]}/{{{N}_{0}}}
	\label{eq:3}
\end{equation}

$h_t$ denotes the ratio of received power to transmitted power in the absence of turbulence, which can be derived from Gaussian beam diffraction propagation as follows:
\begin{equation}
	{{h}_{t}}(\varphi )=\frac{2{{s}_{t}}{{s}_{r}}{{s}_{s}}}{\pi {{R}^{2}}}\frac{1}{{{\omega }^{2}}}\exp \left( -\frac{2{{\varphi }^{2}}}{{{\omega }^{2}}} \right)\cdot \pi {{\left( \frac{{{D}_{r}}}{2} \right)}^{2}}
	\label{eq:4}
\end{equation}

where $R$ is the far-field propagation distance, ${{s}_{t}}$ and ${{s}_{r}}$ are the transmitted and received loss, respectively, ${{s}_{s}}$ is the proportion of split beam for acquisition, ${{D}_{r}}$ is the diameter of the receiver aperture, $\omega $ is divergence angle corresponding to ${1}/{{{e}^{2}}}$ intensity radius of Gaussian beam. $\varphi $ represents the random variable denoting the deviation angle between the transmitter pointing and the LoS under the influence of vibration, as illustrated in Fig. 2. Ref. [13] has previously demonstrated that $\varphi $ follows a Rice distribution. Hence, $E\left[ {{h}_{t}} \right]$ is obtained as:
\begin{equation}
	\begin{aligned}
		& E\left[ {{h}_{t}} \right]=\frac{{{s}_{t}}{{s}_{r}}{{s}_{s}}D_{r}^{2}}{2{{R}^{2}}}{{E}_{\varphi }}\left[ \frac{1}{{{\omega }^{2}}}\exp \left( -\frac{2{{\varphi }^{2}}}{{{\omega }^{2}}} \right) \right] \\ 
		& \ \ \ \ \ \ \ \ \ \ =\frac{{{s}_{t}}{{s}_{r}}{{s}_{s}}D_{r}^{2}}{2{{R}^{2}}}\frac{1}{{{\omega }^{2}}+4{{\sigma }^{2}}}\exp \left( -\frac{2{{\tau }^{2}}{{d}^{2}}}{{{\omega }^{2}}+4{{\sigma }^{2}}} \right) \\ 
	\end{aligned}
	\label{eq:5}
\end{equation}
where $d$ is the distance between adjacent spiral arms, denoted as spiral pitch, $\sigma $ is the standard deviation of the platform vibration. Notably, $0\le \tau \le {1}/{2}$ is coverage factor, representing the ratio of maximum acquisition angle to spiral pitch where the receiver meets lowest SNR level for a certain transmitted power. When the minimum distance between spirals and the receiver exceeds $\tau d$, it implies that the received average SNR consistently falls below the threshold, resulting in acquisition failure.

The atmospheric turbulence is modeled by the Gamma-Gamma distribution\cite{al2001mathematical}. Then the first moment of ${{h}_{c}}$ is given as\cite{wang2010moment}:
\begin{equation}
	E\left[ {{h}_{c}} \right]=\frac{\Gamma (\alpha +k)\Gamma (\beta +1)}{\Gamma (\alpha )\Gamma (\beta )}\left( \frac{\gamma }{\alpha \beta } \right)=\gamma 
	\label{eq:6}
\end{equation}
where $\gamma $ is the scale parameter, $\alpha $ and $\beta $ are large-scale and small-scale effective numbers\cite{prokevs2009modeling}, respectively.

Combining Eqs. (\ref{eq:3}), (\ref{eq:5}) and (\ref{eq:6}), the transmitted power $P_t$ meeting the average SNR threshold at the receiver is derived as:
\begin{equation}
	{{P}_{t}}=B\left( {{\omega }^{2}}+4{{\sigma }^{2}} \right)\exp \left( \frac{2{{\tau }^{2}}{{d}^{2}}}{{{\omega }^{2}}+4{{\sigma }^{2}}} \right)
	\label{eq:7}
\end{equation}
where $B=\frac{\bar{Q}{{N}_{0}}{{R}^{2}}}{{{P}_{L}}R_{r}^{2}{{s}_{t}}{{s}_{r}}{{s}_{s}}\gamma D_{r}^{2}}$.

\section{Single-field Acquisition Energy}
\begin{figure}[t!]
	\centering\includegraphics[width=0.95\columnwidth]{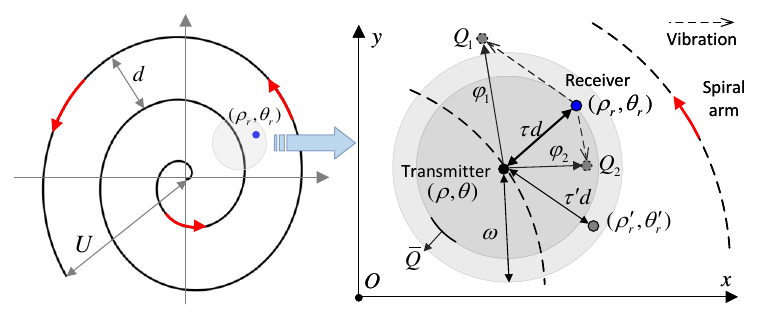}
	\caption{Scanning diagram. $U$ is FOU, $d$ is spiral pitch, and $\omega $ is beam divergence angle. Under the platform vibration, $\varphi $ is the random angle deviation between the transmitter pointing and LoS (the direction pointing towards the receiver). $\mu =\tau d$ is the maximum acquisition angle between spiral arms and receiver meeting the SNR threshold. While ${\mu }'={\tau }'d$ represents an alternative position, but acquisition fails due to ${\mu }'>\mu $.}
	\label{fig:2}
\end{figure}

As depicted in scanning process Fig. \ref{fig:2}, the polar coordinate of initial pointing error (or the position of the receiver) is $({{\rho }_{r}},{{\theta }_{r}})$, obeying the Gaussian distribution with zero mean and equal variance ${{\kappa }_{x}}={{\kappa }_{y}}=\kappa $ in the horizontal and vertical directions. So the polar angle ${{\theta }_{r}}$ follows uniform distribution $U(0,2\pi )$ and the radial error ${{\rho }_{r}}$ obeys Rayleigh distribution as:
\begin{equation}
	{{f}_{{{\rho }_{r}}}}({{\rho }_{r}})=\frac{{{\rho }_{r}}}{{{\kappa }^{2}}}\exp \left( -\frac{\rho _{r}^{2}}{2{{\kappa }^{2}}} \right)
	\label{eq:8}
\end{equation}

The receiver $({{\rho }_{r}},{{\theta }_{r}})$ is likely to be scanned within the FOU, and the corresponding probability is expressed as:
\begin{equation}
	{{P}_{U}}=\int_{0}^{U}{{{f}_{{{\rho }_{r}}}}\left( {{\rho }_{r}} \right)d{{\rho }_{r}}}=1-\exp \left( -\frac{{{U}^{2}}}{2{{\kappa }^{2}}} \right)
	\label{eq:9}
\end{equation}

Given that $\kappa $, measured in milliradian magnitude, is significantly far greater than $d$ in microradian magnitude, the PDF of the coverage factor $\tau $ is approximated as:
\begin{equation}
	\begin{aligned}
		& {{f}_{\tau }}(\tau )\approx \sum\limits_{k=1}^{\infty }{\left[ \exp \left( -\frac{{{(k-1+\tau )}^{2}}{{d}^{2}}}{2{{\kappa }^{2}}} \right)-\exp \left( -\frac{{{(k+\tau )}^{2}}{{d}^{2}}}{2{{\kappa }^{2}}} \right) \right.} \\ 
		& \ \ \ \ \ \ \ \ \ \ \ \ \ \left. -\exp \left( -\frac{{{(k+1-\tau )}^{2}}{{d}^{2}}}{2{{\kappa }^{2}}} \right)+\exp \left( -\frac{{{(k-\tau )}^{2}}{{d}^{2}}}{2{{\kappa }^{2}}} \right) \right] \\ 
		& \ \ \ \ \ \ \ \ =\exp \left( -\frac{{{\tau }^{2}}{{d}^{2}}}{2{{\kappa }^{2}}} \right)+\exp \left( -\frac{{{\left( 1-\tau  \right)}^{2}}{{d}^{2}}}{2{{\kappa }^{2}}} \right)\approx 2 \\ 
	\end{aligned}
	\label{eq:10}
\end{equation}

According to the description in Fig. \ref{fig:2}, The probability that the average received SNR over the threshold is the CDF of the coverage factor:
\begin{equation}
	{{P}_{SNR}}=\int_{0}^{\tau }{{{f}_{\tau }}(\mathcal{T})d\mathcal{T}}=2\tau 
	\label{eq:11}
\end{equation}

In addition, the signal is expected to be detected when the LoS falls within the field angle range of the receiver, and the corresponding field detection probability is denoted as ${{P}_{V}}<1$. We refer to ${{P}_{R}}={{P}_{V}}\cdot {{P}_{SNR}}=2{{P}_{V}}\tau $ as the feedback probability of the receiver. Consequently, the probability of successful acquisition in a single-field scanning is:
\begin{equation}
	{{P}_{S}}={{P}_{U}}\cdot {{P}_{R}}
	\label{eq:12}
\end{equation}

As shown in Fig. \ref{fig:2}, the scanning usually adopts the Archimedean spiral to achieve an efficient search from high probability to low probability regions\cite{steinhaus1999mathematical}. For the radial distance ${{\rho }_{r}}$, the length of the spiral scan is ${\pi \rho _{r}^{2}}/{d}$.

Then the acquisition time is calculated as ${{t}_{S}}={\pi \rho _{r}^{2}}/{(vd)}$, where $v$ is scanning speed. Combined with Eq. (\ref{eq:8}), the PDF of single-field acquisition energy consumption ${{e}_{S}}={{P}_{t}}\cdot {{t}_{S}}$ is obtained as:
\begin{equation}
	{{f}_{{{e}_{S}}}}\left( {{e}_{S}} \right)=\frac{vd}{2\pi {{\kappa }^{2}}{{P}_{t}}}\exp \left( -\frac{vd}{2\pi {{\kappa }^{2}}{{P}_{t}}}{{e}_{S}} \right)
	\label{eq:13}
\end{equation}

Subsequently, when a single-field acquisition is successful, the energy consumption expectation ${{E}_{S}}$ is integrated as:
\begin{equation}
	{{E}_{S}}=\int_{0}^{{{E}_{U}}}{{{e}_{S}}\cdot {{f}_{{{e}_{S}}}}\left( {{e}_{S}} \right)d{{e}_{S}}}=\frac{2\pi {{\kappa }^{2}}{{P}_{t}}}{vd}\left[ 1-\exp \left( -\frac{{{U}^{2}}}{2{{\kappa }^{2}}} \right)\left( 1+\frac{{{U}^{2}}}{2{{\kappa }^{2}}} \right) \right]
	\label{eq:14}
\end{equation}
where ${{E}_{U}}={\pi {{U}^{2}}{{P}_{t}}}/{(vd)}$ represents the energy consumption of scanning the entire FOU.

\section{Multi-field Scanning Model}
Acquisition success cannot be guaranteed with only once single-field scanning due to ${{P}_{S}}<1$. Therefore, the multi-field scanning, which is a series of repetitive single-field scanning over the same FOU, is often employed instead.

\subsection{Scan-stare}
In the case of a single-field scanning failure, it is necessary to reinitialize the transmitter pointing based on ephemeris table. In this process, reset energy ${{E}_{a}}$, which is not related to ${{P}_{t}}$, is strongly essential but ignored by previous analytical models. Therefore, when the acquisition is achieved in $n+1$ single- field, the total scanning energy ${{e}_{M}}$ is:
\begin{equation}
	{{e}_{M}}=n\left( {{E}_{U}}+{{E}_{a}} \right)+{{e}_{S}}
	\label{eq:15}
\end{equation}

The PDF of ${e}_{M}$ is:
\begin{equation}
	{{f}_{{{e}_{M}}}}\left( {{e}_{M}} \right)={{\left( 1-{{P}_{S}} \right)}^{n}}{{P}_{R}}{{f}_{{{e}_{S}}}}\left( {{e}_{S}} \right)
	\label{eq:16}
\end{equation}

The CDF of ${e}_{M}$ is:
\begin{equation}
	\begin{aligned}
		& P\left( E\le {{e}_{M}} \right)=P\left( E\le n\left( {{E}_{U}}+{{E}_{a}} \right) \right)+P\left( n\left( {{E}_{U}}+{{E}_{a}} \right)<E\le {{e}_{M}} \right) \\ 
		& \ \ \ \ \ \ \ \ \ \ \ \ \ \ \ =\sum\limits_{k=0}^{n-1}{{{\left( 1-{{P}_{S}} \right)}^{k}}{{P}_{R}}\int_{0}^{{{E}_{U}}}{{{f}_{{{e}_{S}}}}\left( E \right)dE}}+{{(1-{{P}_{S}})}^{n}}{{P}_{R}}\int_{0}^{{{e}_{S}}}{{{f}_{{{e}_{S}}}}\left( E \right)dE} \\ 
		& \ \ \ \ \ \ \ \ \ \ \ \ \ \ \ =1-{{\left( 1-{{P}_{S}} \right)}^{n}}\left\{ 1-{{P}_{R}}\left[ 1-\exp \left( -\frac{vd}{2\pi {{\kappa }^{2}}{{P}_{t}}}{{e}_{S}} \right) \right] \right\} \\ 
	\end{aligned}
	\label{eq:17}
\end{equation}

There is $n\to \infty $ and ${{e}_{S}}\to {{E}_{U}}$ for ${{e}_{M}}\to \infty $. Then the acquisition probability of Eq. (\ref{eq:17}) becomes:
\begin{equation}
	\underset{{{e}_{M}}\to \infty }{\mathop{\lim }}\,P\left( E\le {{e}_{M}} \right)=\underset{n\to \infty }{\mathop{\lim }}\,\left[ 1-{{\left( 1-{{P}_{S}} \right)}^{n+1}} \right]=1
	\label{eq:18}
\end{equation}

Multi-field scanning can ensure acquisition success from an energy consumption perspective. According to Eq. (\ref{eq:16}), the multi-field acquisition energy expectation ${{E}_{M}}$ is calculated as:
\begin{equation}
	\begin{aligned}
		& {{E}_{M}}=\int_{0}^{\infty }{{{e}_{M}}\cdot {{f}_{{{e}_{M}}}}\left( {{e}_{M}} \right)d{{e}_{M}}}=\sum\limits_{n=0}^{\infty }{{{\left( 1-{{P}_{S}} \right)}^{n}}{{P}_{R}}\int_{0}^{{{E}_{U}}}{\left[ n\left( {{E}_{U}}+{{E}_{a}} \right)+{{e}_{S}} \right]\cdot {{f}_{{{e}_{S}}}}\left( {{e}_{S}} \right)d{{e}_{S}}}} \\ 
		& \ \ \ \ =\sum\limits_{n=0}^{\infty }{{{\left( 1-{{P}_{S}} \right)}^{n}}{{P}_{S}}\left[ \frac{{{E}_{S}}}{{{P}_{U}}}+n\left( {{E}_{U}}+{{E}_{a}} \right) \right]}=\frac{{{E}_{S}}}{{{P}_{U}}}+\left( \frac{1}{{{P}_{S}}}-1 \right)\left( {{E}_{U}}+{{E}_{a}} \right) \\ 
		& \ \ \ \ =\frac{2\pi {{\kappa }^{2}}{{P}_{t}}}{vd}\left[ \frac{{{e}^{\eta }}\eta \left( 1-{{P}_{R}} \right)}{\left( {{e}^{\eta }}-1 \right){{P}_{R}}}+1 \right]+\frac{{{E}_{a}}\left[ {{e}^{\eta }}\left( 1-{{P}_{R}} \right)+{{P}_{R}} \right]}{\left( {{e}^{\eta }}-1 \right){{P}_{R}}} \\ 
	\end{aligned}
	\label{eq:19}
\end{equation}
where $\eta ={{{U}^{2}}}/{(2{{\kappa }^{2}})}>0$. Obviously, ${{E}_{M}}$ is a function of the parameters $(d,\tau ,\omega ,U)$, which are optimized respectively to minimize the acquisition energy.

\subsubsection{Optimizations}
The partial derivative of ${{E}_{M}}$ with respect to $d$ is:
\begin{equation}
	\frac{\partial {{E}_{M}}}{\partial d}=\frac{2\pi {{\kappa }^{2}}{{P}_{t}}}{v{{d}^{2}}\left( {{\omega }^{2}}+4{{\sigma }^{2}} \right)}\left[ \frac{{{e}^{\eta }}\eta \left( 1-{{P}_{R}} \right)}{\left( {{e}^{\eta }}-1 \right){{P}_{R}}}+1 \right]\left( 4{{\tau }^{2}}{{d}^{2}}-{{\omega }^{2}}-4{{\sigma }^{2}} \right)
	\label{eq:20}
\end{equation}
where the minimum is taken at ${\partial {{E}_{M}}}/{\partial d=0}$, with the optimum ${{d}_{opt}}$ as:
\begin{equation}
	{{d}_{opt}}=\frac{\sqrt{{{\omega }^{2}}+4{{\sigma }^{2}}}}{2\tau }
	\label{eq:21}
\end{equation}

Substituting it into Eq. (\ref{eq:19}) yields:
\begin{equation}
	{{\left. {{E}_{M}} \right|}_{d={{d}_{opt}}}}=\frac{2\pi {{e}^{{1}/{2}\;}}{{\kappa }^{2}}B}{v{{\left( {{\omega }^{2}}+4{{\sigma }^{2}} \right)}^{{-1}/{2}\;}}}\left[ \frac{{{e}^{\eta }}\eta \left( 1-{{P}_{R}} \right)}{\left( {{e}^{\eta }}-1 \right){{P}_{V}}}+\frac{{{P}_{R}}}{{{P}_{V}}} \right]+\frac{{{E}_{a}}\left[ {{e}^{\eta }}\left( 1-{{P}_{R}} \right)+{{P}_{R}} \right]}{\left( {{e}^{\eta }}-1 \right){{P}_{R}}}
	\label{eq:22}
\end{equation}

Then take the partial derivative concerning $\omega $:
\begin{equation}
	\frac{\partial {{E}_{M}}}{\partial \omega }=\frac{2\pi {{e}^{{1}/{2}\;}}{{\kappa }^{2}}B\omega }{v\sqrt{{{\omega }^{2}}+4{{\sigma }^{2}}}}\left[ \frac{{{e}^{\eta }}\eta \left( 1-{{P}_{R}} \right)}{\left( {{e}^{\eta }}-1 \right){{P}_{V}}}+\frac{{{P}_{R}}}{{{P}_{V}}} \right]
	\label{eq:23}
\end{equation}
which is always greater than zero, so the optimum beam divergence angle ${{\omega }_{opt}}$ is the lower limit of the divergence angle ${{\omega }_{limit}}$ that the laser can modulate.

Next, the partial derivative of Eq. (\ref{eq:22}) with respect to $\tau $ is:
\begin{equation}
	\frac{\partial {{E}_{M}}}{\partial \tau }=-\frac{4\pi {{e}^{{1}/{2}\;}}{{\kappa }^{2}}B\left( {{e}^{\eta }}\eta -{{e}^{\eta }}+1 \right)}{v{{\left( {{\omega }^{2}}+4{{\sigma }^{2}} \right)}^{{-1}/{2}\;}}\left( {{e}^{\eta }}-1 \right)}-\frac{{{E}_{a}}{{e}^{\eta }}}{\left( {{e}^{\eta }}-1 \right){{P}_{R}}\tau }
	\label{eq:24}
\end{equation}
where ${{e}^{\eta }}\eta -{{e}^{\eta }}+1$ is positive constantly. Thus ${{E}_{M}}$ decreases monotonically with $\tau $, and the minimum ${{E}_{M}}$ is reached at ${{\tau }_{opt}}={1}/{2}$. 

Furthermore, the partial derivative of ${{E}_{M}}$ concerning $\eta $ is calculated after taking ${{\tau }_{opt}}$ into Eq. (\ref{eq:22}):
\begin{equation}
	\frac{\partial {{E}_{M}}}{\partial \eta }=\frac{2\pi {{e}^{{1}/{2}\;}}{{\kappa }^{2}}B\left( 1-{{P}_{V}} \right){{e}^{\eta }}\left( {{e}^{\eta }}-\eta -1-{{{\hat{E}}}_{a}} \right)}{v{{\left( {{\omega }^{2}}+4{{\sigma }^{2}} \right)}^{{-1}/{2}\;}}{{\left( {{e}^{\eta }}-1 \right)}^{2}}{{P}_{V}}}
	\label{eq:25}
\end{equation}
where ${{\hat{E}}_{a}}=\frac{v{{\left( {{\omega }^{2}}+4{{\sigma }^{2}} \right)}^{{-1}/{2}\;}}{{E}_{a}}}{2\pi {{e}^{{1}/{2}\;}}{{\kappa }^{2}}B\left( 1-{{P}_{V}} \right)}$. The minimum ${{E}_{M}}$ is taken at ${\partial {{E}_{M}}}/{\partial \eta }=0$:
\begin{equation}
	{{e}^{\eta }}-\eta -1-{{\hat{E}}_{a}}=0
	\label{eq:26}
\end{equation}

However, the above equation has no analytical solution. When $\eta $ is large, there is approximate $\eta \approx \ln ({{\hat{E}}_{a}})$, which can be employed as the variable to perform polynomial fitting with numerical solutions and effectively reduce the order. In general, $\omega $ is ${{10}^{-5}}$ magnitude, $v$ and $\kappa $ are of the same order of ${{10}^{-3}}$ magnitude, $(1-{{P}_{V}})$ and ${{E}_{a}}$ are the level of ${{10}^{-2}}$ and ${{10}^{2}}$, respectively, and $B$ is at ${{10}^{7}}$ level. Therefore, ${{\hat{E}}_{a}}$ is about ${{10}^{-1}}\sim{{10}^{0}}$ magnitude order. Without loss of generality, we adopt polynomials to fit the numerical solutions, where the goodness of fit (GoF) is utilized as an index to evaluate the fitting accuracy. We perform piecewise fitting in the interval $\left[ 0.01,10 \right]$ for $x=\ln ({{\hat{E}}_{a}})$ with $GoF=0.999$:
\begin{equation}
	{{\eta }_{opt}}=\left\{ \begin{aligned}
		& 0.02785{{x}^{2}}+0.3123x+0.9870,\ \ 0.01\le {{{\hat{E}}}_{a}}<0.1 \\ 
		& 0.06049{{x}^{2}}+0.4552x+1.1454,\ \ \ \ 0.1\le {{{\hat{E}}}_{a}}\le 1 \\ 
		& 0.07181{{x}^{2}}+0.4713x+1.1455,\ \ \ \ \ \ \ 1<{{{\hat{E}}}_{a}}\le 10 \\ 
	\end{aligned} \right.
	\label{eq:27}
\end{equation}

Consequently, the optimum FOU is ${U}_{opt}=\kappa \sqrt{2{{\eta }_{opt}}}$.

\subsection{Dual-scan}
In this section, we investigate another type of acquisition process known as the dual-scan. This mode involves two terminals simultaneously performing scan. Once one of the terminals acquires the other, both terminals stop scanning and proceed to the pointing operation. Therefore, the dual-scan process can be visualized as two individual scan-stare procedures.

We assume that the parameters of both terminals are the same. Two independent random variables ${{E}_{1}}$ and ${{E}_{2}}$ represent the energy required for T1 to acquire T2 and T2 to acquire T1, respectively. The energy consumption of acquisition first is lower, and the random variable for dual-scan acquisition energy consumption is represented by\cite{hechenblaikner2021analysis}:
\begin{equation}
	E=2\cdot \min \left\{ {{E}_{1}},{{E}_{2}} \right\}
	\label{eq:28}
\end{equation}

For the energy consumption ${{\tilde{e}}_{M}}=n\left( {{{\tilde{E}}}_{U}}+{{{\tilde{E}}}_{a}} \right)+{{\tilde{e}}_{S}}$, the CDF is computed as:
\begin{equation}
	\begin{aligned}
		& {{F}_{\min }}\left( {{\tilde{e}}}_{M} \right)={{P}_{\min }}\left( E\le {{{\tilde{e}}}_{M}} \right)=1-{{P}_{\min }}\left( E>{{{\tilde{e}}}_{M}} \right)=1-P\left( {{E}_{1}}>\frac{{{\tilde{e}}}_{M}}{2} \right)\cdot P\left( {{E}_{2}}>\frac{{{\tilde{e}}}_{M}}{2} \right) \\ 
		& \ \ \ \ \ \ \ \ \ \ \ \ \ =1-{{\left( 1-{{P}_{S}} \right)}^{2n}}{{\left\{ 1-{{P}_{R}}\left[ 1-\exp \left( -\frac{vd\left[ {{{\tilde{e}}}_{M}}-n\left( {{{\tilde{E}}}_{U}}+{{{\tilde{E}}}_{a}} \right) \right]}{2\pi {{\kappa }^{2}}{{{\tilde{P}}}_{t}}} \right) \right] \right\}}^{2}} \\ 
	\end{aligned}
	\label{eq:29}
\end{equation}
where ${{\tilde{E}}_{U}}=2{{E}_{U}}$, ${{\tilde{E}}_{a}}=2{{E}_{a}}$, and ${{\tilde{P}}_{t}}=2{{P}_{t}}$. Then we get the corresponding PDF by taking the derivative concerning ${{\tilde{e}}_{M}}$:
\begin{equation}
	{{f}_{{{{\tilde{e}}}_{M}}}}\left( {{{\tilde{e}}}_{M}} \right)={{\left( 1-{{{\tilde{P}}}_{S}} \right)}^{n}}{{P}_{R}}{{f}_{{{{\tilde{e}}}_{S}}}}\left( {{{\tilde{e}}}_{S}} \right)
	\label{eq:30}
\end{equation}
where ${{\tilde{P}}_{S}}={{P}_{S}}\left( 2-{{P}_{S}} \right)$. Eq. (\ref{eq:30}) shares a similar expression form with Eq. (\ref{eq:16}). Consequently, we are able to transform the complex dual-scan into the simpler scan-stare, which has been previously analyzed albeit with distinct parameters. The corresponding PDF of single-field acquisition energy is generalized as:
\begin{equation}
	{{f}_{{{{\tilde{e}}}_{S}}}}\left( {{{\tilde{e}}}_{S}} \right)=\frac{vd}{\pi {{\kappa }^{2}}{{{\tilde{P}}}_{t}}}\exp \left( -\frac{vd{{{\tilde{e}}}_{S}}}{\pi {{\kappa }^{2}}{{{\tilde{P}}}_{t}}} \right)\left[ \exp \left( \frac{vd{{{\tilde{e}}}_{S}}}{2\pi {{\kappa }^{2}}{{{\tilde{P}}}_{t}}} \right)\left( 1-{{P}_{R}} \right)+{{P}_{R}} \right]
	\label{eq:31}
\end{equation}

Analogously, the probability ${{\tilde{P}}_{U}}$ of falling within the FOU is:
\begin{equation}
	{{\tilde{P}}_{U}}=\int_{0}^{{{{\tilde{E}}}_{U}}}{{{f}_{{{{\tilde{e}}}_{S}}}}\left( {{{\tilde{e}}}_{S}} \right)d{{{\tilde{e}}}_{S}}}=\left( 1-{{e}^{-\eta }} \right)\left( 2-{{P}_{R}}+{{P}_{R}}{{e}^{-\eta }} \right)
	\label{eq:32}
\end{equation}

Then the single-field acquisition energy ${{\tilde{E}}_{S}}$ is obtained as:
\begin{equation}
	\begin{aligned}
		& {{{\tilde{E}}}_{S}}=\int_{0}^{{{{\tilde{E}}}_{U}}}{{{{\tilde{e}}}_{S}}\cdot {{f}_{{{{\tilde{e}}}_{S}}}}\left( {{{\tilde{e}}}_{S}} \right)d{{{\tilde{e}}}_{S}}} \\ 
		& \ \ \ =\frac{\pi {{\kappa }^{2}}{{{\tilde{P}}}_{t}}\left[ {{e}^{2\eta }}\left( 4-3{{P}_{R}} \right)-4{{e}^{\eta }}\left( \eta +1 \right)\left( 1-{{P}_{R}} \right)-{{P}_{R}}\left( 2\eta +1 \right) \right]}{vd\cdot {{e}^{2\eta }}} \\ 
	\end{aligned}
	\label{eq:33}
\end{equation}

Consequently, the multi-field acquisition energy expectation ${{\tilde{E}}_{M}}$ is calculated:
\begin{equation}
	\begin{aligned}
		& {{{\tilde{E}}}_{M}}=\frac{{{{\tilde{E}}}_{S}}}{{{{\tilde{P}}}_{U}}}+\left( \frac{1}{{{{\tilde{P}}}_{S}}}-1 \right)\left( {{{\tilde{E}}}_{U}}+{{{\tilde{E}}}_{a}} \right) \\ 
		& \ \ \ \ \ =\frac{{{{\tilde{E}}}_{a}}vd{{\left[ {{e}^{\eta }}\left( 1-{{P}_{R}} \right)+{{P}_{R}} \right]}^{2}}+\pi {{\kappa }^{2}}{{{\tilde{P}}}_{t}}\cdot H\left( \eta ,{{P}_{R}} \right)}{vd\left( {{e}^{\eta }}-1 \right){{P}_{R}}\left[ {{e}^{\eta }}\left( 2-{{P}_{R}} \right)+{{P}_{R}} \right]} \\ 
	\end{aligned}
	\label{eq:34}
\end{equation}

\begin{equation}
	H\left( \eta ,{{P}_{R}} \right)={{e}^{2\eta }}\left[ {{P}_{R}}\left( 4-3{{P}_{R}} \right)+2\eta {{\left( 1-{{P}_{R}} \right)}^{2}} \right]-4{{e}^{\eta }}{{P}_{R}}\left( 1-{{P}_{R}} \right)-P_{R}^{2}
	\label{eq:35}
\end{equation}

\subsubsection{Optimizations}
The partial derivative of ${{\tilde{E}}_{M}}$ with respect to $d$ is:
\begin{equation}
	\frac{\partial {{{\tilde{E}}}_{M}}}{\partial d}=\frac{\pi {{\kappa }^{2}}{{{\tilde{P}}}_{t}}\left( 4{{\tau }^{2}}{{d}^{2}}-{{\omega }^{2}}-4{{\sigma }^{2}} \right)\cdot H\left( \eta ,{{P}_{R}} \right)}{v{{d}^{2}}\left( {{\omega }^{2}}+4{{\sigma }^{2}} \right)\left( {{e}^{\eta }}-1 \right){{P}_{R}}\left[ {{e}^{\eta }}\left( 2-{{P}_{R}} \right)+{{P}_{R}} \right]}
	\label{eq:36}
\end{equation}
where $H\left( \eta ,{{P}_{R}} \right)$ of Eq. (\ref{eq:35}) is rearranged as a quadratic equation of ${{P}_{R}}$:
\begin{equation}
	H\left( \eta ,{{P}_{R}} \right)={{\mathcal{H}}_{1}}\left( \eta  \right)P_{R}^{2}-{{\mathcal{H}}_{2}}\left( \eta  \right){{P}_{R}}+{{\mathcal{H}}_{3}}\left( \eta  \right)
	\label{eq:37}
\end{equation}
where ${{\mathcal{H}}_{1}}\left( \eta  \right)=2{{e}^{2\eta }}\left( \eta -1 \right)+2{{e}^{\eta }}-{{({{e}^{\eta }}-1)}^{2}}>0$, ${{\mathcal{H}}_{2}}\left( \eta  \right)=4{{e}^{2\eta }}\left( \eta -1 \right)+4{{e}^{\eta }}>0$, and ${{\mathcal{H}}_{3}}\left( \eta  \right)=2\eta \cdot {{e}^{2\eta }}$. The axis of symmetry is $\frac{{{\mathcal{H}}_{2}}\left( \eta  \right)}{2{{\mathcal{H}}_{1}}\left( \eta  \right)}>1$, obtaining:
\begin{equation}
	H\left( \eta ,{{P}_{R}} \right)\ge H\left( \eta ,1 \right)={{e}^{2\eta }}-1>0
	\label{eq:38}
\end{equation}

Therefore, the optimum spiral pitch ${{d}_{opt}}$ is same with Eq. (\ref{eq:21}). Analogously, we bring it into Eq. (\ref{eq:34}) and then get:
\begin{equation}
	{{\left. {{{\tilde{E}}}_{M}} \right|}_{d={{d}_{opt}}}}\ =\frac{2\pi {{e}^{{1}/{2}\;}}{{\kappa }^{2}}B\sqrt{{{\omega }^{2}}+4{{\sigma }^{2}}}\cdot H\left( \eta ,{{P}_{R}} \right)}{v\left( {{e}^{\eta }}-1 \right){{P}_{V}}\left[ {{e}^{\eta }}\left( 2-{{P}_{R}} \right)+{{P}_{R}} \right]}+\frac{{{{\tilde{E}}}_{a}}{{\left[ {{e}^{\eta }}\left( 1-{{P}_{R}} \right)+{{P}_{R}} \right]}^{2}}}{\left( {{e}^{\eta }}-1 \right){{P}_{R}}\left[ {{e}^{\eta }}\left( 2-{{P}_{R}} \right)+{{P}_{R}} \right]}
	\label{eq:39}
\end{equation}

The partial derivative of Eq. (\ref{eq:39}) concerning $\omega $:
\begin{equation}
	\frac{\partial {{{\tilde{E}}}_{M}}}{\partial \omega }=\frac{\omega }{\sqrt{{{\omega }^{2}}+4{{\sigma }^{2}}}}\cdot \frac{2\pi {{e}^{{1}/{2}\;}}{{\kappa }^{2}}B\cdot H\left( \eta ,{{P}_{R}} \right)}{v\left( {{e}^{\eta }}-1 \right){{P}_{V}}\left[ {{e}^{\eta }}\left( 2-{{P}_{R}} \right)+{{P}_{R}} \right]}
	\label{eq:40}
\end{equation}

It can be obtained from Eq. (\ref{eq:38}) that ${\partial {{{\tilde{E}}}_{M}}}/{\partial \omega }$ greater than zero constantly. Thereby the optimum beam divergence angle ${{\omega }_{opt}}$ is also equal ${{\omega }_{limit}}$.

Meanwhile the partial derivative with respect to $\tau $ is:
\begin{equation}
	\frac{\partial {{{\tilde{E}}}_{M}}}{\partial \tau }=\frac{4\pi {{e}^{{1}/{2}\;}}{{\kappa }^{2}}B\sqrt{{{\omega }^{2}}+4{{\sigma }^{2}}}P_{R}^{2}\cdot J\left( \eta ,{{P}_{R}} \right)-4v{{{\tilde{E}}}_{a}}{{e}^{2\eta }}{{P}_{V}}\left[ {{e}^{\eta }}\left( 1-{{P}_{R}} \right)+{{P}_{R}} \right]}{v\left( {{e}^{\eta }}-1 \right)P_{R}^{2}{{\left[ {{e}^{\eta }}\left( 2-{{P}_{R}} \right)+{{P}_{R}} \right]}^{2}}}
	\label{eq:41}
\end{equation}
\begin{equation}
	\begin{aligned}
		& J\left( \eta ,{{P}_{R}} \right)={{e}^{3\eta }}\left[ \left( 3-2\eta  \right){{\left( 2-{{P}_{R}} \right)}^{2}}+2\eta -4 \right]+{{e}^{\eta }}\left( 5{{P}_{R}}-4 \right){{P}_{R}} \\ 
		& \ \ \ \ \ \ \ \ \ \ \ \ \ +{{e}^{2\eta }}\left[ P_{R}^{2}\left( 2\eta -7 \right)+16{{P}_{R}}-2\eta -8 \right]-P_{R}^{2} \\ 
	\end{aligned}
	\label{eq:42}
\end{equation}

In addition, the partial derivative of $J\left( \eta ,{{P}_{R}} \right)$ concerning ${{P}_{R}}$ is:
\begin{equation}
	\frac{\partial J\left( \eta ,{{P}_{R}} \right)}{\partial {{P}_{R}}}=\left[ {{e}^{\eta }}\left( 2-{{P}_{R}} \right)+{{P}_{R}} \right]\left[ {{e}^{2\eta }}\left( 2\eta -3 \right)+4{{e}^{\eta }}-1 \right]>0
	\label{eq:43}
\end{equation}

It is shown that $J(\eta, P_R)$ is monotonically increasing with respect to $P_R$, with its maximum value achieved at $P_R = 1$, i.e., $J\left( \eta ,1 \right)=-\left( {{e}^{2\eta }}-1 \right)\left( {{e}^{\eta }}-1 \right)<0$. Therefore, $J(\eta,P_R)$ is always less than zero, and ${\partial {{{\tilde{E}}}_{M}}}/{\partial \tau }$ as well. Consequently, the minimum $\tilde{E}_M$ is achieved at ${{\tau }_{opt}}={1}/{2}$.

Furthermore, the partial derivative concerning $\eta $ is calculated after taking ${{\tau }_{opt}}={1}/{2}$ into Eq. (\ref{eq:39}) as:
\begin{equation}
	\frac{\partial {{{\tilde{E}}}_{M}}}{\partial \eta }=\frac{4\pi {{e}^{{1}/{2}\;}}{{\kappa }^{2}}B\left( 1-{{P}_{V}} \right){{e}^{\eta }}\left[ {{e}^{\eta }}\left( 1-{{P}_{V}} \right)+{{P}_{V}} \right]\cdot G\left( \eta ,{{P}_{V}},{{{\hat{E}}}_{a}} \right)}{{{\left( {{\omega }^{2}}+4{{\sigma }^{2}} \right)}^{{-1}/{2}\;}}v{{\left( {{e}^{\eta }}-1 \right)}^{2}}{{P}_{V}}{{\left[ {{e}^{\eta }}\left( 2-{{P}_{V}} \right)+{{P}_{V}} \right]}^{2}}}
	\label{eq:44}
\end{equation}
\begin{equation}
	G\left( \eta ,{{P}_{V}},{{{\hat{E}}}_{a}} \right)={{e}^{2\eta }}\left( 2-{{P}_{V}} \right)-2{{e}^{\eta }}\left( \eta -{{P}_{V}}\eta +{{{\hat{E}}}_{a}}+1 \right)+{{P}_{V}}
	\label{eq:45}
\end{equation}

The partial derivative of the key term $G\left( \eta ,{{P}_{V}},{{{\hat{E}}}_{a}} \right)$ with respect to $\eta $ is:
\begin{equation}
	\frac{\partial G\left( \eta ,{{P}_{V}},{{{\hat{E}}}_{a}} \right)}{\partial \eta }=2{{e}^{\eta }}\left[ \left( 2-{{P}_{V}} \right)\left( {{e}^{\eta }}-\eta -1 \right)+\eta -{{{\hat{E}}}_{a}} \right]
	\label{eq:46}
\end{equation}

It can be seen that as $\eta$ increases, $G\left( \eta ,{{P}_{V}},{{{\hat{E}}}_{a}} \right)$ first decreases and then increases. Since $G\left( 0,{{P}_{V}},{{{\hat{E}}}_{a}} \right)=-2{{\hat{E}}_{a}}<0$ and $G({{\hat{E}}_{a}}+1,{{P}_{V}},{{\hat{E}}_{a}})>(2-{{P}_{V}}){{e}^{\eta }}({{e}^{\eta }}-2\eta )>0$, $G(\eta, P_V, \hat{E}_a)$ must have a unique zero point corresponding to the minimum $\tilde{E}_M$. However, it is hard to solve $G\left( \eta ,{{P}_{V}},{{{\hat{E}}}_{a}} \right)=0$. We employ $x=\ln(\hat{E}_a)$ and $P_V$ as variables and perform polynomials fitting with numerical solutions, with $\hat{E}_a$ and $P_V$ within $\left[0.01, 10\right]$ and $\left[0.1, 1\right)$, respectively, as well as $GoF=0.999$. The piecewise fitted polynomials are:
\begin{equation}
	{{\eta }_{opt}}=\left\{ \begin{aligned}
		& 0.0305{{x}^{2}}+0.0115{{P}_{V}}x+0.002P_{V}^{2}\ \ ,0.01\le {{{\hat{E}}}_{a}}<0.1 \\ 
		& +0.3279x+0.0487{{P}_{V}}+1.01 \\ 
		& 0.0701{{x}^{2}}+0.0685{{P}_{V}}x+0.0318P_{V}^{2}\ \ ,0.1\le {{{\hat{E}}}_{a}}\le 1 \\ 
		& +0.4678x+0.1261{{P}_{V}}+1.147 \\ 
		& 0.0721{{x}^{2}}+0.1384{{P}_{V}}x+0.1485P_{V}^{2}\ \ \ \ ,1<{{{\hat{E}}}_{a}}\le 10 \\ 
		& +0.4602x+0.0057{{P}_{V}}+1.168 \\ 
	\end{aligned} \right.
	\label{eq:47}
\end{equation}

Thereby ${{U}_{opt}}=\kappa \sqrt{2{{\eta }_{opt}}}$.

\section{Discussion of the results}
\begin{figure}[t!]
	\centering\includegraphics[width=1\columnwidth]{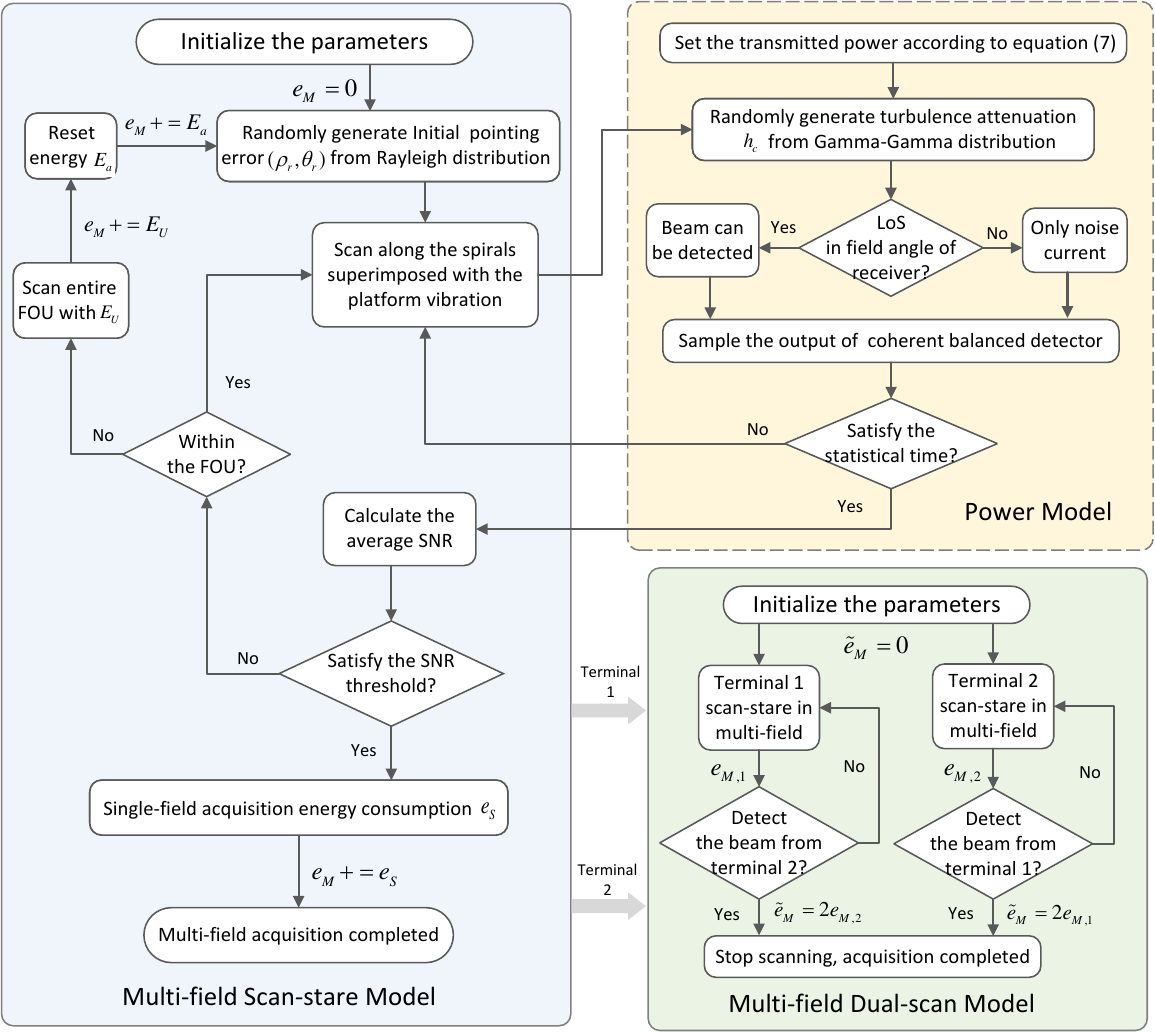}
	\caption{Simulation process of Monto Carlo}
	\label{fig:3}
\end{figure}
In this section, MC simulations are conducted to validate the derived analytical expressions. The simulation time step aligns with the sampling interval of the photoelectric sensor. Following the methodology outlined in Ref.\cite{kasdin1995discrete}, random amplitudes and phases are assigned based on the power spectral density of the platform vibration, yielding a frequency-domain sequence. This sequence undergoes inverse Fourier transformation to generate a random time-domain sequence representing platform vibration. This sequence is then superimposed onto the nominal spirals to obtain the simulated scanning trajectory. The simulation process is depicted in Fig. \ref{fig:3}, accompanied by the corresponding simulation parameters listed in Table \ref{tab:1}.
\begin{table}[t!]
	\centering
	\caption{Simulation parameters}
	\label{tab:1}
	\begin{tabular}{@{}lll@{}}
		\toprule
		Parameters                        & Value & Unit   \\ \midrule
		Link distance $R$                    & 36000 & $km$     \\
		Transmitter loss $s_t$                  & 0.92  & -      \\
		Receiver loss $s_r$                     & 0.92  & -      \\
		Receiver aperture diameter $D_r$            & 30 & $cm$    \\
		Proportion of split beam $s_s$				& 0.1 & - \\
		Photoelectric response efficiency $R_r$  & 0.88   & -      \\
		Std. of noise current ${\sigma }_n$             & 8   & $nA$ \\
		Local oscillator power $P_L$             & 0.1   & $mW$ \\
		Receiver average SNR threshold $\bar{Q}$   & 10    & $dB$     \\
		Statistical time for average SNR		 &	1	& $s$		\\
		Band-limited platform vibration			 & 100  &  $Hz$		\\
		Sample frequency of photodetector & 10   &  $KHz$		\\
		Weak turbulence scale $\gamma$	& 0.9 & - \\
		Medium turbulence scale $\gamma$	& 0.5 & - \\
		Std. of platform vibration $\sigma $        & 10     & $\mu rad$  \\
		Std. of initial LOS error $\kappa$         & 1  & $mrad$       \\
		Scanning speed $v$                    & 10  & $\mu rad/s$       \\
		Reset energy ${E}_{a}$                  & 1  & $J$       \\
		Field detection probability ${P}_{V}$  & 0.95  & -        \\ \bottomrule
	\end{tabular}
\end{table}

\begin{figure}[t!]
	\centering\includegraphics[width=1\columnwidth]{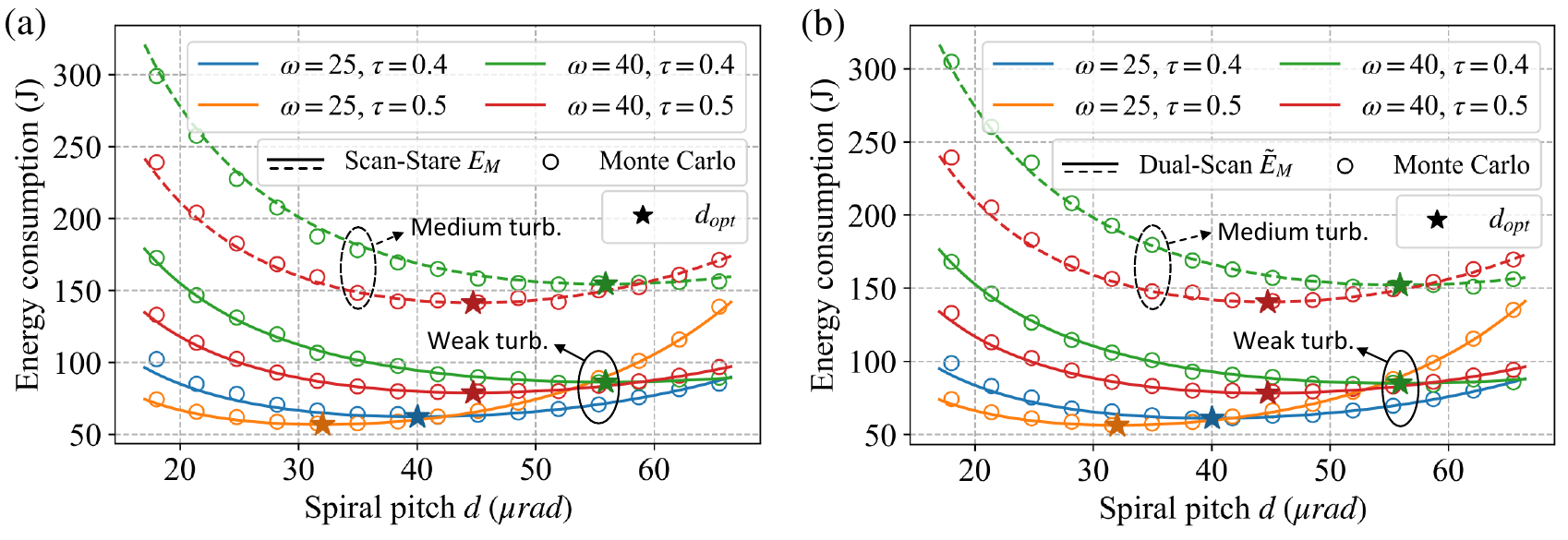}
	\caption{The variation of the multi-field acquisition energy with spiral pitch under different combinations of turbulence, beam divergence angle, and coverage factor. (a) scan-stare and (b) dual-scan.}
	\label{fig:4}
\end{figure}
Fig. \ref{fig:4} shows the variation of the multi-field acquisition energy of (a) scan-stare and (b) dual-scan modes with spiral pitch under different combinations of turbulence, beam divergence angle, and coverage factor, where ${U}/{\kappa }=1.3$. The simulation process is illustrated in Fig. \ref{fig:3}. The theoretical values ${{E}_{M}}$ and ${{\tilde{E}}_{M}}$ are calculated according to Eqs. (\ref{eq:19}) and (\ref{eq:34}), and the optimum spiral pitch ${d}_{opt}$ is obtained from Eq. (\ref{eq:21}). It can be observed that the theoretical results are consistent with the corresponding MC results. When ${d}<{{d}_{opt}}$, the scan time is longer, and while ${d}>{{d}_{opt}}$, the transmitted power that satisfies the received SNR threshold is larger, both of which lead to excessive acquisition energy. In addition, as the beam divergence angle increases and the coverage factor decreases, ${d}_{opt}$ gradually increases. Meanwhile, the increase of turbulence only increases the acquisition energy without changing ${d}_{opt}$. Consequently, the optimization conclusion regarding the spiral pitch is validated. It is observed that the energy of both modes is nearly identical, and relevant quantitative analyses are subsequently conducted.

\begin{figure}[t!]
	\centering\includegraphics[width=1\columnwidth]{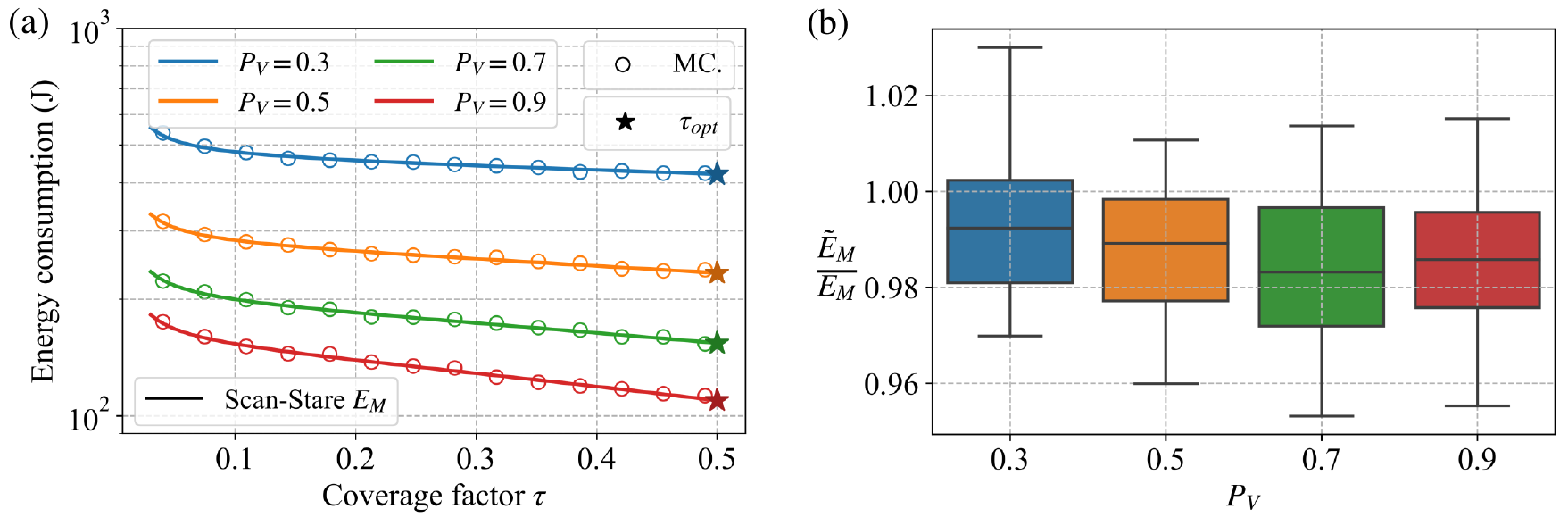}
	\caption{The results under different field detection probabilities vary with the coverage factor. (a) Multi-field acquisition energy for scan-stare. (b) Ratio of energy consumption between dual-scan and scan-stare modes.}
	\label{fig:5}
\end{figure}

Fig. \ref{fig:5} shows the variation of (a) acquisition energy and (b) ratio of multi-field scan-stare and dual-scan modes with coverage factor under different field detection probabilities, respectively, where $\omega =25\mu rad$, ${U}/{\kappa }=1.3$, spiral pitch is calculated from Eq. (\ref{eq:21}), and the turbulence is medium. The acquisition energy decreases with the increase of coverage factor, and this trend is also reflected in Fig. \ref{fig:4}. In addition, an increase in field detection probability ${P_V}$ means a greater single-field acquisition probability, which can effectively reduce the number of scans without changing the single-field scanning time and reduce the acquisition energy. Therefore, it is advisable to select a photodetector with a larger field angle as much as possible. Observing box plot (b), it is apparent that the median of the ratio is below one, suggesting that the acquisition energy consumption of the dual-scan is statistically lower than that of the scan-stare mode.

\begin{figure}[t!]
	\centering\includegraphics[width=1\columnwidth]{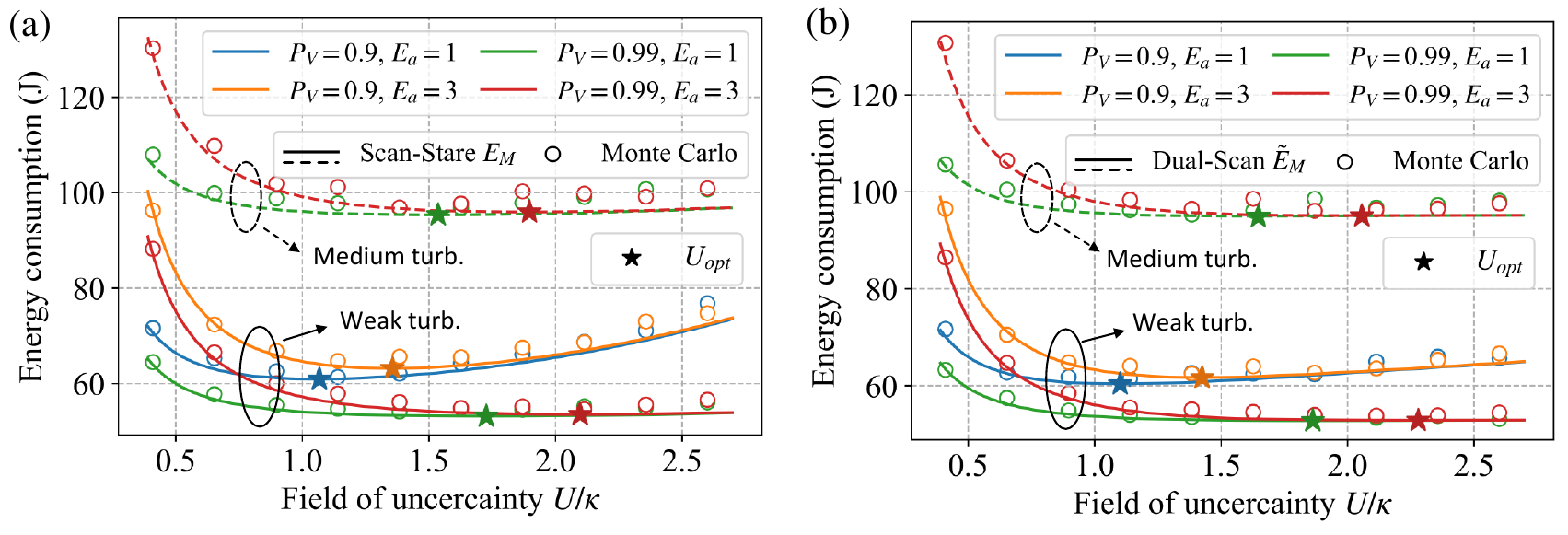}
	\caption{The variation of multi-field acquisition energy with FOU under different combinations of turbulence, field detection probability, and reset energy. (a) scan-stare. (b) dual-scan.}
	\label{fig:6}
\end{figure}

Fig. \ref{fig:6} shows the variation of multi-field acquisition energy of (a) scan-stare and (b) dual-scan with FOU under different combinations of turbulence, field detection probability, and reset energy, respectively, where $\omega =25\mu rad$, $\tau = {1}/{2}$, and the spiral pitch is calculated from Eq. (\ref{eq:21}) as $d=32\mu rad$. The theoretical optimum FOU ${{U}_{opt}}$ for scan-stare and dual-scan modes are fitted by Eq. (\ref{eq:27}) and (\ref{eq:47}), respectively. The theoretical results are observed to align well with the corresponding Monte Carlo (MC) results. For small values of $U$, variations in reset energy significantly affect acquisition energy consumption. At${U}/{\kappa }=0.4$, the acquisition energy consumption of ${{E}_{a}}=3$ increases by 38\% compared to that of ${{E}_{a}}=1$. Conversely, for large values such as ${U}/{\kappa }=2.5$, the acquisition energy consumption of ${{E}_{a}}=3$ increases by only 0.1\% compared to that of ${{E}_{a}}=1$. In this case, the impact of reset energy variation can be neglected, and scanning energy consumption plays a dominant role. This is because with the increase of FOU, the single-field acquisition probability increases, and the scanning time decrease, leading to a decrease in total reset energy. On the other hand, the energy for scanning the entire FOU ${{E}_{U}}$ or ${{{\tilde{E}}}_{U}}$ is proportional to the square of $U$, which is a high-order term compared to ${{E}_{a}}$. In addition, when $U>{{U}_{opt}}$, the acquisition energy almost does not change with the variation of FOU in the case of ${{P}_{V}} = 0.99$. Theoretically, when ${{P}_{R}} = {{P}_{SNR}}{{P}_{V}} = 0.99\to 1$, the acquisition energy for scan-stare and dual-scan are ${{E}_{M}}\to {2\pi {{\kappa }^{2}}{{P}_{t}}}/{(vd)}$ and ${{\tilde{E}}_{M}}\to {\pi {{\kappa }^{2}}{{{\tilde{P}}}_{t}}}/{(vd)}$, respectively, which are not related to FOU. This means that the selection of FOU can be more flexible. Meanwhile, increasing ${{E}_{a}}$ and ${{P}_{V}}$ or decreasing turbulence leads to an increase in ${{U}_{opt}}$. Furthermore, the difference between the fitted ${{U}_{opt}}$ and the corresponding acquisition energy of scan-stare and dual-scan with the ground-truth values are within 1\% and ${ 10 }^{-3}\%$ as well as within 1\% and ${ 10 }^{-2}\%$, respectively, validating the optimization conclusion about FOU.

\begin{figure}[t!]
	\centering\includegraphics[width=1\columnwidth]{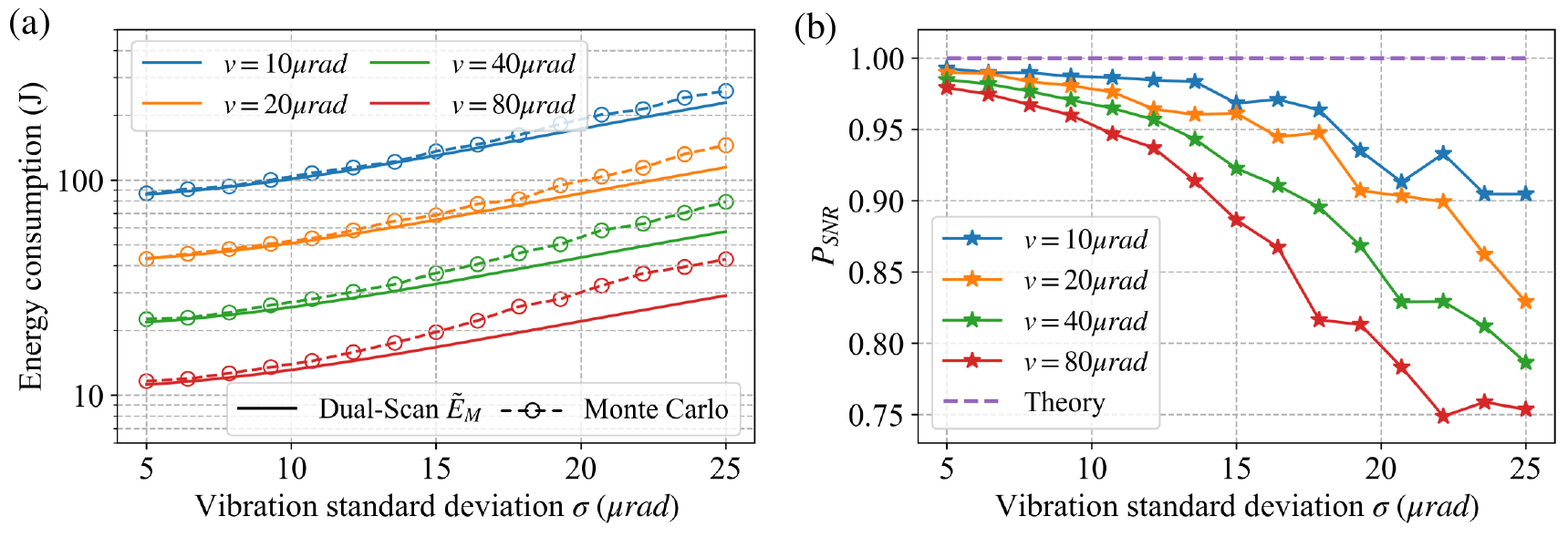}
	\caption{The results of dual-scan under different scanning speeds as the platform vibration increases. (a) Multi-field acquisition energy consumption plotted on a logarithmic scale. (b) Probability of average SNR exceeding the threshold.}
	\label{fig:7}
\end{figure}

Through the above quantitative analysis, it can be inferred that under the same parameters, the multi-field acquisition energy consumption of both modes is almost identical. Consequently, Fig. \ref{fig:7} examines the variation of platform vibration at different scanning speeds in the more complex dual-scan mode, where $\omega =25\mu rad$, $d=30\mu rad$, $\tau ={1}/{2}$, ${U}/{\kappa }=1.3$, and the turbulence is medium. To maintain a consistent scan length for average SNR statistics, the duration of the statistical time is set as follows: 1 second for $v=10\mu rad$, 0.5 seconds for $v=20\mu rad$, 0.25 seconds for $v=40\mu rad$, and 0.125 seconds for $v=80\mu rad$. In Fig. \ref{fig:7} (a), it is evident that as the platform vibration intensity increases, the acquisition energy consumption gradually rises. This is because the power required at the transmitter increases to meet the SNR threshold at the receiver. Moreover, maintaining the same scan length at different speeds ensures consistent expectations for statistically averaged SNR. However, increasing the vibration intensity and reducing the SNR statistical time lead to increased variance, resulting in a flatter probability density curve. Consequently, this results in a decrease in the probability of the average SNR exceeding the threshold, as depicted in Fig. \ref{fig:7} (b), leading to a decline in single-field acquisition probability and causing the simulated energy consumption to deviate further from the theoretical value illustrated in Fig. \ref{fig:7} (a). Notably, the consistency between simulation and theory is highest at $v=10\mu rad$.

\begin{figure}[t!]
	\centering\includegraphics[width=1\columnwidth]{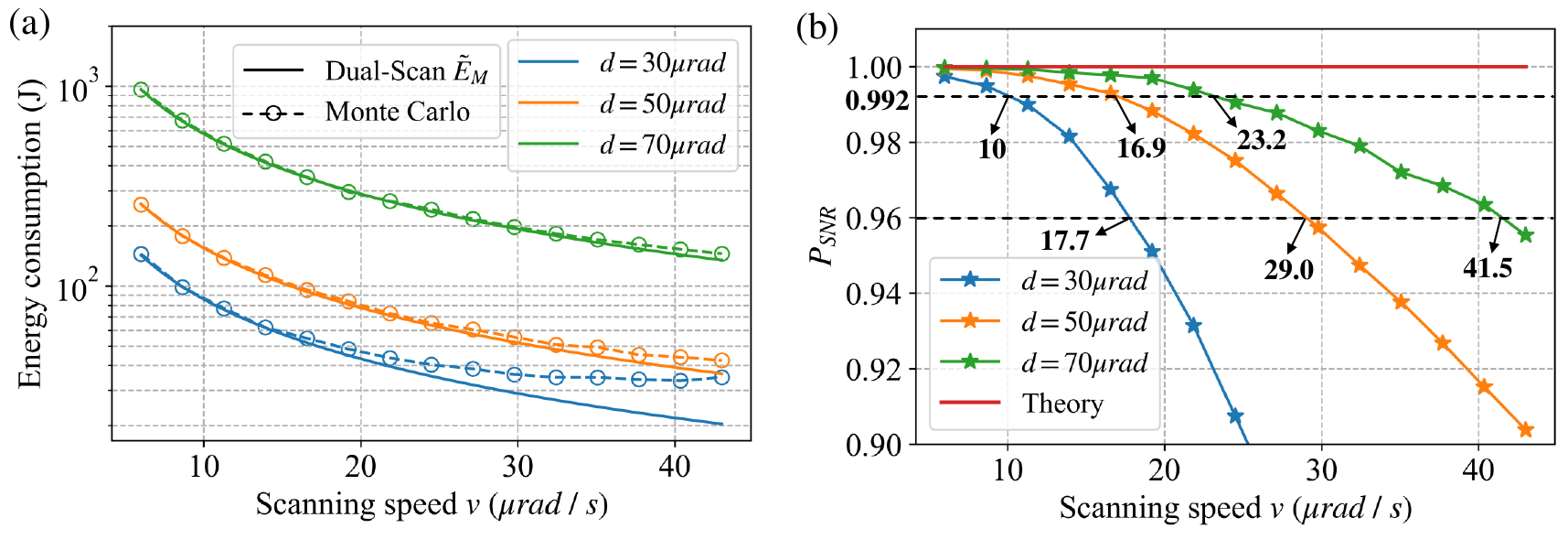}
	\caption{The results for dual-scan under different spiral pitches as the scanning speed changes. (a) Multi-field acquisition energy consumption plotted on a logarithmic scale. (b) Probability of average SNR over the threshold.}
	\label{fig:8}
\end{figure}

Subsequently, Fig. \ref{fig:8} presents simulation on the variation of scanning speed under different spiral pitches, where $\omega =25\mu rad$, $\tau ={1}/{2}$, ${U}/{\kappa }=1.3$, and the turbulence is medium. In Fig. \ref{fig:8} (a), as the scanning speed increases, the scanning time decreases, leading to a reduction in multi-field acquisition energy consumption. Additionally, the simulated energy consumption gradually deviates from the theoretical value as the scanning speed increases. This is because the increased scanning length within the 1-second period for average SNR statistics causes some trajectory points to be significantly farther from the receiver, resulting in a significant decrease in received power and consequently reducing the average SNR at the receiver, as shown in Fig. \ref{fig:8} (b), causing a decrease in single-field acquisition probability. Moreover, in Fig. \ref{fig:8} (b), the ${{P}_{SNR}}$ located on the curve of $d=30\mu rad$ corresponding to $v=10\mu rad$ is 0.992, which is annotated with a black dashed line. Subsequently, the speeds on the curves for $d=50\mu rad$ and $d=70\mu rad$, where the consistency between simulation and theory is 99.2\%, are recorded as $16.9{\mu rad}/{s}$ and $23.2{\mu rad}/{s}$, respectively. Given that the maximum acquisition angles $\tau d$ satisfying the SNR threshold for three spiral pitches are $15\mu rad$, $25\mu rad$, and $35\mu rad$, respectively, the ratios of scanning length within the 1-second statistical time to maximum acquisition angle (denoted as $\varepsilon $) are calculated as 0.67, 0.68, and 0.66, respectively, indicating their close proximity. To validate this quantitative relationship, the speeds where the consistency between simulation and theory is 96\% on three spiral pitches are identified as $17.7{\mu rad}/{s}$, $29.0{\mu rad}/{s}$, and $41.5{\mu rad}/{s}$, respectively, then the corresponding $\varepsilon $ are 1.18, 1.16, and 1.19, respectively, demonstrating their close proximity. Furthermore, the simulation consistencies corresponding to different $\varepsilon $ are listed in Table \ref{tab:2}. It is apparent that to hold good simulation consistency, the selection of $\varepsilon $ should be less than one.

\begin{table}[t!]
	\centering
	\caption{The Relationship between $\varepsilon $ and Simulation Consistency}
	\label{tab:2}
	\begin{tabular}{@{}cc@{}}
		\toprule
		The ratio $\varepsilon $ & Simulation Consistency (\%)  \\ \midrule
		0.5                                                   & 99.63                       \\
		1                                                     & 97.64                       \\
		1.5                                                   & 92.6                        \\
		2                                                     & 85.3                        \\
		2.5                                                   & 75.88                       \\
		3                                                     & 63.86                       \\ \bottomrule
		\multicolumn{2}{l}{\begin{tabular}[c]{@{}l@{}}\small $\varepsilon $ is the ratio of scan length within the time for average \\
				\small SNR statistics to maximum acquisition angle.\end{tabular}}
	\end{tabular}
\end{table}

\begin{figure}[t!]
	\centering\includegraphics[width=0.65\columnwidth]{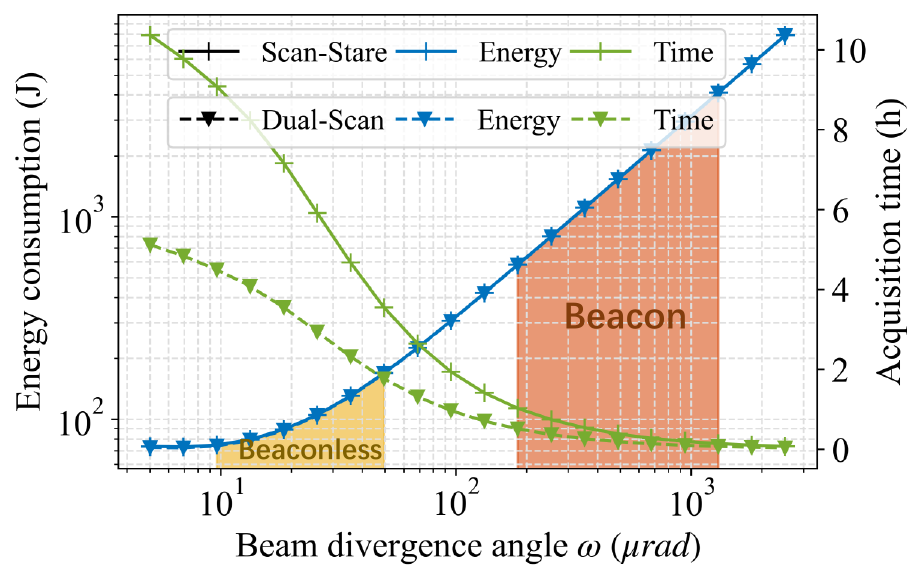}
	\caption{The variations of acquisition energy consumption and time with the beam divergence angle. The beaconless and the beacon are represented with orange and red regions, respectively.}
	\label{fig:9}
\end{figure}

Finally, Fig. \ref{fig:9} describes the changes in multi-field acquisition energy consumption for two modes with increasing beam divergence angle, where the turbulence is medium, ${U}/{\kappa }=1.3$, $d=30\mu rad$, $\tau ={1}/{2}$, and $v={10\mu rad}/{s}$ to hold 99\% simulation consistency. The corresponding acquisition time is also plotted in Fig. \ref{fig:9}. It is evident that the energy curves of the two modes almost overlap. Moreover, as $\omega$ increases, although the acquisition time decreases, the energy increases. This is because the transmitted power required increases significantly to guarantee that the received average SNR is greater than the threshold. In particular, the typical divergence angle of the beaconless is about $10\sim50\mu rad$, highlighted in the orange region, and the average acquisition energy is $107J$ . While the divergence angle of the beacon is typically $10\sim30$ times greater than the beaconless, shown as the red region, and the average acquisition energy is $1881J$, which is about 18 times greater than that of the beaconless. Furthermore, the minimum energy consumption is achieved at minimum beam divergence angle ${\omega }_{limit}$, which is also the divergence angle during the APT tracking phase. This quantitatively concludes that the beaconless is superior to the beacon for APT with the goal of minimizing energy consumption. Given that the acquisition time of the dual-scan is about half that of the scan-stare, thus the former outperforms the latter in terms of acquisition efficiency.

\section{Conclusion}
This paper conducts a detailed and comprehensive study of energy consumption, an important resource for the satellite to execute missions, based on the received average SNR with multi-field scanning for GEO-to-ground optical link acquisition. We first establish a link model from transmitted power to received SNR under the coherent detection system, involving factors such as turbulence, platform vibration, beam divergence angle, spiral pitch, coverage factor, as well as various losses. This model is applicable to various photodetectors, thus holding practical guidance significance. Combined with the acquisition time and the essential reset energy preparing for the next single-field scanning, we derive the closed-form expression of multi-field acquisition energy and perform optimization in the scan-stare mode. Subsequently, we discuss another mode called the dual-scan, where two terminals synchronously scan until one acquires the other. Through simulations, we find that the acquisition energy of the dual-scan is lower than that of scan-stare. Meanwhile, the acquisition time of the former is about half of the latter, which further enhances the advantage of dual-scan in acquisition efficiency. Moreover, simulation results regarding platform vibration and scanning speed variations indicate that when the ratio of scan length within the time for average SNR statistics to maximum acquisition angle is less than one, the consistency between the theoretical model and simulation results exceeds 97\%. Notably, the proposed model can be extended to optical link acquisition scenarios between satellites in the same orbit due to the consideration of turbulence effects. Finally, we compare the acquisition energy of the beaconless and the beacon in their typical range, and observe that the energy consumption with beaconless is only 6\% of that with beacon, indicating that the beaconless is a better strategy for optical link acquisition with the goal of energy optimization, providing strong theoretical support for FSOC system miniaturization.

\section*{Disclosures}
The authors declare no conflicts of interest.

\bibliographystyle{unsrtnat}
\bibliography{references}  

\begin{thebibliography}{37}
\providecommand{\natexlab}[1]{#1}
\providecommand{\url}[1]{\texttt{#1}}
\expandafter\ifx\csname urlstyle\endcsname\relax
  \providecommand{\doi}[1]{doi: #1}\else
  \providecommand{\doi}{doi: \begingroup \urlstyle{rm}\Url}\fi

\bibitem[Toyoshima et~al.(2007)Toyoshima, Leeb, Kunimori, and
  Takano]{toyoshima2007comparison}
Morio Toyoshima, Walter~R Leeb, Hiroo Kunimori, and Tadashi Takano.
\newblock Comparison of microwave and light wave communication systems in space
  applications.
\newblock \emph{Optical engineering}, 46\penalty0 (1):\penalty0 015003--015003,
  2007.

\bibitem[Toyoshima(2005)]{toyoshima2005trends}
Morio Toyoshima.
\newblock Trends in satellite communications and the role of optical free-space
  communications.
\newblock \emph{Journal of Optical Networking}, 4\penalty0 (6):\penalty0
  300--311, 2005.

\bibitem[Kim et~al.(2001)Kim, Riley, Wong, Mitchell, Brown, Hakakha, Adhikari,
  and Korevaar]{kim2001lessons}
Isaac~I Kim, Brian Riley, Nicholas~M Wong, Mary Mitchell, Wesley Brown, Harel
  Hakakha, Prasanna Adhikari, and Eric~J Korevaar.
\newblock Lessons learned for strv-2 satellite-to-ground lasercom experiment.
\newblock In \emph{Free-Space Laser Communication Technologies XIII}, volume
  4272, pages 1--15. SPIE, 2001.

\bibitem[Fields et~al.(2011)Fields, Kozlowski, Yura, Wong, Wicker, Lunde,
  Gregory, Wandernoth, and Heine]{fields20115}
Renny Fields, David Kozlowski, Harold Yura, Robert Wong, Josef Wicker, C~Lunde,
  Mark Gregory, B~Wandernoth, and F~Heine.
\newblock 5.625 gbps bidirectional laser communications measurements between
  the nfire satellite and an optical ground station.
\newblock In \emph{2011 International Conference on Space Optical Systems and
  Applications (ICSOS)}, pages 44--53. IEEE, 2011.

\bibitem[Benzi et~al.(2016)Benzi, Troendle, Shurmer, James, Lutzer, and
  Kuhlmann]{benzi2016optical}
Edoardo Benzi, Daniel~C Troendle, Ian Shurmer, Mark James, Michael Lutzer, and
  Sven Kuhlmann.
\newblock Optical inter-satellite communication: the alphasat and sentinel-1a
  in-orbit experience.
\newblock In \emph{14th International Conference on Space Operations}, page
  2389, 2016.

\bibitem[Hauschildt et~al.(2019)Hauschildt, le~Gallou, Mezzasoma, Moeller,
  Armengol, Witting, Herrmann, and Carmona]{hauschildt2019global}
Harald Hauschildt, Nicolas le~Gallou, Silvia Mezzasoma, Hermann~Ludwig Moeller,
  Josep~Perdigues Armengol, Michael Witting, J{\"o}rg Herrmann, and Cesar
  Carmona.
\newblock Global quasi-real-time-services back to europe: Edrs global.
\newblock In \emph{International Conference on Space Optics—ICSO 2018},
  volume 11180, pages 353--357. SPIE, 2019.

\bibitem[Young et~al.(1986)Young, Germann, and Nelson]{young1986pointing}
Philip~W Young, Lawrence~M Germann, and Roy Nelson.
\newblock Pointing, acquisition, and tracking subsystem for space-based laser
  communications.
\newblock In \emph{Optical technologies for communication satellite
  applications}, volume 616, pages 118--128. SPIE, 1986.

\bibitem[Wuchenich et~al.(2014)Wuchenich, Mahrdt, Sheard, Francis, Spero,
  Miller, Mow-Lowry, Ward, Klipstein, Heinzel, et~al.]{wuchenich2014laser}
Danielle~MR Wuchenich, Christoph Mahrdt, Benjamin~S Sheard, Samuel~P Francis,
  Robert~E Spero, John Miller, Conor~M Mow-Lowry, Robert~L Ward, William~M
  Klipstein, Gerhard Heinzel, et~al.
\newblock Laser link acquisition demonstration for the grace follow-on mission.
\newblock \emph{Optics express}, 22\penalty0 (9):\penalty0 11351--11366, 2014.

\bibitem[Barausse et~al.(2020)Barausse, Berti, Hertog, Hughes, Jetzer, Pani,
  Sotiriou, Tamanini, Witek, Yagi, et~al.]{barausse2020prospects}
Enrico Barausse, Emanuele Berti, Thomas Hertog, Scott~A Hughes, Philippe
  Jetzer, Paolo Pani, Thomas~P Sotiriou, Nicola Tamanini, Helvi Witek, Kent
  Yagi, et~al.
\newblock Prospects for fundamental physics with lisa.
\newblock \emph{General Relativity and Gravitation}, 52:\penalty0 1--33, 2020.

\bibitem[Luo et~al.(2021)Luo, Wang, Wu, Hu, and Jin]{luo2021taiji}
Ziren Luo, Yan Wang, Yueliang Wu, Wenrui Hu, and Gang Jin.
\newblock The taiji program: A concise overview.
\newblock \emph{Progress of Theoretical and Experimental Physics},
  2021\penalty0 (5):\penalty0 05A108, 2021.

\bibitem[Guelman et~al.(2004)Guelman, Kogan, Kazarian, Livne, Orenstein, and
  Michalik]{guelman2004acquisition}
M~Guelman, Anton Kogan, A~Kazarian, A~Livne, M~Orenstein, and H~Michalik.
\newblock Acquisition and pointing control for inter-satellite laser
  communications.
\newblock \emph{IEEE Transactions on Aerospace and Electronic Systems},
  40\penalty0 (4):\penalty0 1239--1248, 2004.

\bibitem[Toyoshima et~al.(2002)Toyoshima, Jono, Nakagawa, and
  Yamamoto]{toyoshima2002optimum}
Morio Toyoshima, Takashi Jono, Keizo Nakagawa, and Akio Yamamoto.
\newblock Optimum divergence angle of a gaussian beam wave in the presence of
  random jitter in free-space laser communication systems.
\newblock \emph{JOSA A}, 19\penalty0 (3):\penalty0 567--571, 2002.

\bibitem[Steinhaus(1999)]{steinhaus1999mathematical}
Hugo Steinhaus.
\newblock \emph{Mathematical snapshots}.
\newblock Courier Corporation, 1999.

\bibitem[Yang and Li(2022{\natexlab{a}})]{yang2022derivation}
Sen Yang and Xiaofeng Li.
\newblock Derivation of ambiguity in wavefront aberration and quantitative
  analysis in ao system.
\newblock \emph{Optics and Lasers in Engineering}, 158:\penalty0 107174,
  2022{\natexlab{a}}.

\bibitem[Yang and Li(2022{\natexlab{b}})]{yang2022iterative}
Sen Yang and Xiaofeng Li.
\newblock Iterative framework for a high accuracy aberration estimation with
  one-shot wavefront sensing.
\newblock \emph{Optics Express}, 30\penalty0 (21):\penalty0 37874--37887,
  2022{\natexlab{b}}.

\bibitem[Fu et~al.(2021)Fu, Chen, Guan, Zhang, Wu, and Wang]{fu2021virtual}
Wujie Fu, Shaokang Chen, Jian Guan, Zhongtan Zhang, Hao Wu, and Dong~F Wang.
\newblock A virtual-movement scheme for eliminating spot-positioning errors
  applicable to quadrant detectors.
\newblock \emph{IEEE Transactions on Instrumentation and Measurement},
  70:\penalty0 1--11, 2021.

\bibitem[Qiu et~al.(2021)Qiu, Lin, and Chen]{qiu2021active}
Zhaobing Qiu, Liyu Lin, and Liqiong Chen.
\newblock An active method to improve the measurement accuracy of four-quadrant
  detector.
\newblock \emph{Optics and Lasers in Engineering}, 146:\penalty0 106718, 2021.

\bibitem[Yu et~al.(2017)Yu, Wu, Wang, Tan, and Ma]{yu2017theoretical}
Siyuan Yu, Feng Wu, Qiang Wang, Liying Tan, and Jing Ma.
\newblock Theoretical analysis and experimental study of constraint boundary
  conditions for acquiring the beacon in satellite--ground laser
  communications.
\newblock \emph{Optics Communications}, 402:\penalty0 585--592, 2017.

\bibitem[Hu et~al.(2022)Hu, Yu, Duan, Zhu, Cao, Zhou, Li, and Liu]{hu2022multi}
Siqi Hu, Hanghua Yu, Zheng Duan, Ye~Zhu, Caixia Cao, Miaomiao Zhou, Guotong Li,
  and Huijie Liu.
\newblock Multi-parameter influenced acquisition model with an in-orbit jitter
  for inter-satellite laser communication of the lces system.
\newblock \emph{Optics Express}, 30\penalty0 (19):\penalty0 34362--34377, 2022.

\bibitem[Hindman and Robertson(2004)]{hindman2004beaconless}
Charles Hindman and Lawrence Robertson.
\newblock Beaconless satellite laser acquisition-modeling and feasability.
\newblock In \emph{IEEE MILCOM 2004. Military Communications Conference,
  2004.}, volume~1, pages 41--47. IEEE, 2004.

\bibitem[Ho(2007)]{ho2007pointing}
Tzung-Hsien Ho.
\newblock \emph{Pointing, acquisition, and tracking systems for free-space
  optical communication links}.
\newblock University of Maryland, College Park, 2007.

\bibitem[Li et~al.(2011)Li, Yu, Ma, and Tan]{li2011analytical}
Xin Li, Siyuan Yu, Jing Ma, and Liying Tan.
\newblock Analytical expression and optimization of spatial acquisition for
  intersatellite optical communications.
\newblock \emph{Optics Express}, 19\penalty0 (3):\penalty0 2381--2390, 2011.

\bibitem[Friederichs et~al.(2016)Friederichs, Sterr, and
  Dallmann]{friederichs2016vibration}
Lothar Friederichs, Uwe Sterr, and Daniel Dallmann.
\newblock Vibration influence on hit probability during beaconless spatial
  acquisition.
\newblock \emph{Journal of Lightwave Technology}, 34\penalty0 (10):\penalty0
  2500--2509, 2016.

\bibitem[Ma et~al.(2021)Ma, Lu, Tan, Yu, Fu, and Li]{ma2021satellite}
Jing Ma, Gaoyuan Lu, Liying Tan, Siyuan Yu, Yulong Fu, and Fajun Li.
\newblock Satellite platform vibration influence on acquisition system for
  intersatellite optical communications.
\newblock \emph{Optics \& Laser Technology}, 138:\penalty0 106874, 2021.

\bibitem[Hechenblaikner(2021)]{hechenblaikner2021analysis}
Gerald Hechenblaikner.
\newblock Analysis of performance and robustness against jitter of various
  search methods for acquiring optical links in space.
\newblock \emph{Applied Optics}, 60\penalty0 (13):\penalty0 3936--3946, 2021.

\bibitem[Kaushal and Kaddoum(2016)]{kaushal2016optical}
Hemani Kaushal and Georges Kaddoum.
\newblock Optical communication in space: Challenges and mitigation techniques.
\newblock \emph{IEEE communications surveys \& tutorials}, 19\penalty0
  (1):\penalty0 57--96, 2016.

\bibitem[Li(2015)]{li2015limited}
Hanshan Li.
\newblock Limited magnitude calculation method and optics detection performance
  in a photoelectric tracking system.
\newblock \emph{Applied Optics}, 54\penalty0 (7):\penalty0 1612--1617, 2015.

\bibitem[Chin et~al.(2018)Chin, Brandon, Bugga, Smart, Jones, Krause, West, and
  Bolotin]{chin2018energy}
Keith~B Chin, Erik~J Brandon, Ratnakumar~V Bugga, Marshall~C Smart, Simon~C
  Jones, Frederick~C Krause, William~C West, and Gary~G Bolotin.
\newblock Energy storage technologies for small satellite applications.
\newblock \emph{Proceedings of the IEEE}, 106\penalty0 (3):\penalty0 419--428,
  2018.

\bibitem[Cui et~al.(2023)Cui, Zhu, Xiao, Clerckx, and Zhang]{cui2023energy}
Huanxi Cui, Lipeng Zhu, Zhenyu Xiao, Bruno Clerckx, and Rui Zhang.
\newblock Energy-efficient rsma for multigroup multicast and multibeam
  satellite communications.
\newblock \emph{IEEE Wireless Communications Letters}, 2023.

\bibitem[Andrews and Phillips(2005)]{andrews2005laser}
Larry~C Andrews and Ronald~L Phillips.
\newblock Laser beam propagation through random media.
\newblock \emph{Laser Beam Propagation Through Random Media: Second Edition},
  2005.

\bibitem[Popoola et~al.(2012)Popoola, Poves, and Haas]{popoola2012spatial}
Wasiu~O Popoola, Enrique Poves, and Harald Haas.
\newblock Spatial pulse position modulation for optical communications.
\newblock \emph{Journal of Lightwave Technology}, 30\penalty0 (18):\penalty0
  2948--2954, 2012.

\bibitem[Peppas and Mathiopoulos(2015)]{peppas2015free}
Kostas~P Peppas and P~Takis Mathiopoulos.
\newblock Free-space optical communication with spatial modulation and coherent
  detection over hk atmospheric turbulence channels.
\newblock \emph{Journal of Lightwave Technology}, 33\penalty0 (20):\penalty0
  4221--4232, 2015.

\bibitem[Jurado-Navas et~al.(2012)Jurado-Navas, Garrido-Balsells, Paris,
  Castillo-V{\'a}zquez, and Puerta-Notario]{jurado2012impact}
Antonio Jurado-Navas, Jos{\'e}~Mar{\'\i}a Garrido-Balsells, Jos{\'e}~Francisco
  Paris, Miguel Castillo-V{\'a}zquez, and Antonio Puerta-Notario.
\newblock Impact of pointing errors on the performance of generalized
  atmospheric optical channels.
\newblock \emph{Optics Express}, 20\penalty0 (11):\penalty0 12550--12562, 2012.

\bibitem[Al-Habash et~al.(2001)Al-Habash, Andrews, and
  Phillips]{al2001mathematical}
Ammar Al-Habash, Larry~C Andrews, and Ronald~L Phillips.
\newblock Mathematical model for the irradiance probability density function of
  a laser beam propagating through turbulent media.
\newblock \emph{Optical engineering}, 40\penalty0 (8):\penalty0 1554--1562,
  2001.

\bibitem[Wang and Cheng(2010)]{wang2010moment}
Ning Wang and Julian Cheng.
\newblock Moment-based estimation for the shape parameters of the gamma-gamma
  atmospheric turbulence model.
\newblock \emph{Optics express}, 18\penalty0 (12):\penalty0 12824--12831, 2010.

\bibitem[Proke{\v{s}}(2009)]{prokevs2009modeling}
Ale{\v{s}} Proke{\v{s}}.
\newblock Modeling of atmospheric turbulence effect on terrestrial fso link.
\newblock \emph{Radioengineering}, 18\penalty0 (1):\penalty0 42--47, 2009.

\bibitem[Kasdin(1995)]{kasdin1995discrete}
N~Jeremy Kasdin.
\newblock Discrete simulation of colored noise and stochastic processes and
  1/f/sup/spl alpha//power law noise generation.
\newblock \emph{Proceedings of the IEEE}, 83\penalty0 (5):\penalty0 802--827,
  1995.

\end{thebibliography}






\end{document}